\documentclass[journal,onecolumn]{IEEEtran}

\usepackage{amsmath}
\usepackage{colordvi}
\usepackage{amsfonts}
\usepackage{graphicx}
\usepackage{epsfig}
\usepackage{color}
\usepackage{multirow}
\usepackage{url}
\usepackage{bm}

\def\eps{\varepsilon}
\def\P{\mathbb{P}}
\def\S{{\cal S}}

\def\et{\textit{et al.}}

\def\conv{*}

\def\eps{\varepsilon}

\def\S{{\cal S}}

\def\N{{\mathbb{N}}}

\def\Ar{A^{\textrm{r}}}
\def\Zr{Z^{\textrm{r}}}

\def\S{\textbf{S}}

\newtheorem{theorem}{Theorem}

\newtheorem{lemma}{Lemma}
\newtheorem{definition}{Definition}

 {\begin{list}{}%
         {\setlength{\leftmargin}{#1}}%
         \item[]%
 }
 {\end{list}}

\begin{document}

\title{Sharp Bounds in Stochastic Network Calculus}

\author{Florin Ciucu, Felix Poloczek, and Jens Schmitt

\thanks{A 2-page extended abstract of this paper appeared at ACM Sigmetrics 2013.}
\thanks{Florin Ciucu and Felix Poloczek are with Telekom Innovation Laboratories / TU Berlin, Germany, e-mails: \{florin,felix\}@net.t-labs.tu-berlin.de.}
\thanks{Jens Schmitt is with the Department of Computer Science, University of Kaiserslautern, Germany, e-mail: jschmitt@informatik.uni-kl.de}
}
\maketitle

\begin{abstract}
The practicality of the stochastic network calculus (SNC) is often
questioned on grounds of potential looseness of its performance
bounds. In this paper it is uncovered that for bursty arrival
processes (specifically Markov-Modulated On-Off (MMOO)), whose
amenability to \textit{per-flow} analysis is typically proclaimed
as a highlight of SNC, the bounds can unfortunately indeed be very
loose (e.g., by several orders of magnitude off). In response to
this uncovered weakness of SNC, the (Standard) per-flow bounds are
herein improved by deriving a general sample-path bound, using
martingale based techniques, which accommodates FIFO, SP, EDF, and
GPS scheduling. The obtained (Martingale) bounds gain an
exponential decay factor of ${\mathcal{O}}\left(e^{-\alpha
n}\right)$ in the number of flows $n$. Moreover, numerical
comparisons against simulations show that the Martingale bounds
are remarkably accurate for FIFO, SP, and EDF scheduling; for GPS
scheduling, although the Martingale bounds substantially improve
the Standard bounds, they are numerically loose, demanding for
improvements in the core SNC analysis of GPS.
\end{abstract}

\section{Introduction}
Several approaches to the classical queueing theory have emerged
over the past decades. For instance, matrix analytic methods (MAM)
not only provide a unified treatment for a large class of queueing
systems, but they also lend themselves to practical numerical
solutions; two key ideas are the proper accounting of the
repetitive structure of underlying Markov processes, and the use
of linear algebra rather than classic methods based on real
analysis (see~Neuts~\cite{Neuts1981book} and
Lipsky~\cite{Lipsky08}). Another unified approach targeting broad
classes of queueing problems is the stochastic network calculus
(SNC) (see~Chang~\cite{Book-Chang} and Jiang and
Liu~\cite{Jiang08}), which can be regarded as a mixture between
the deterministic network calculus conceived by Cruz~\cite{Cruz91}
(see Le Boudec and Thiran~\cite{Book-LeBoudec}) and the effective
bandwidth theory (see Kelly~\cite{Kelly96}). Because SNC solves
queueing problems in terms of~\textit{bounds}, it is often
regarded as an unconventional approach, especially by the queueing
theory community.

MAM and SNC could be (slightly) compared by the way they apply to
queues with fluid arrivals. In their simplest form, fluid arrival
models were defined as Markov-Modulated On-Off (MMOO) processes by
Anick, Mitra, and Sondhi~\cite{ANIC82}, and were significantly
extended thereafter, especially for the purpose of modelling the
increasingly prevalent voice and video traffic in the Internet. By
relating fluid models and Quasi-Birth-Death (QBD) processes,
Ramaswami has argued that MAM can lend themselves to numerically
more accurate solutions than spectral analysis
methods~\cite{Ramaswami99}. In turn, SNC can produce alternative
solutions with negligible numerical complexity, but these are
arguably less relevant than exact solutions (simply because they
are expressed as bounds). What does, therefore, justify more than
two decades of research in SNC?

The answer lies in two key features of SNC: \textit{scheduling
abstraction} and \textit{convolution-form networks}~(see Ciucu and
Schmitt~\cite{Ciucu12}). The former expresses the ability of SNC
to compute per-flow (or per-class) queueing metrics for a large
class of scheduling algorithms, and in a unified manner.
Concretely, given a flow $A$ sharing a queueing system with other
flows, the characteristics of the scheduling algorithm are first
abstracted away in the so-called \textit{service process};
thereafter, the derivation of queueing metrics for the flow $A$ is
scheduling independent. Furthermore, the per-flow results can be
extended in a straightforward manner from a single queue to a
large class of queueing networks (typically feed-forward), which
are amenable to a convolution-form representation in an
appropriate algebra.

By relying on these two features, SNC could tackle several open
queueing networks problems. The typical scenario involves the
computation of end-to-end (e2e) non-asymptotic performance bounds
(e.g., on the delay distribution) of a single flow crossing a
tandem network and sharing the single queues with some other
flows. Such scenarios were solved for a large class of arrival
processes (see, e.g., Ciucu~\et~\cite{CiBuLi06,BuLiPa06} and
Fidler~\cite{Fidler06} for MMOO processes, and
Liebeherr~\et~\cite{LiBuCi12} for heavy-tailed and self-similar
processes). Another important solution was given for the e2e delay
distribution in a tandem (packet) network with Poisson arrival and
exponential packet sizes, by circumventing Kleinrock's
independence assumption on the regeneration of packet sizes at
each node (see~Burchard~\et~\cite{BuLiCi11}). Other fundamentally
difficult problems include the performance analysis of stochastic
networks implementing network coding (see Yuan~\et~\cite{Wu10}),
the delay analysis of wireless channels under Markovian
assumptions (see Zheng~\et~\cite{Zheng13}), the delay analysis of
multi-hop fading channels (see Al-Zubaidy~\et~\cite{Zubaidy13}),
or bridging information theory and queueing theory by accounting
for the stochastic nature and delay-sensitivity of real sources
(see L{\"u}bben and Fidler~\cite{Lubben11}).

Based on its ability to partially solve fundamentally hard
queueing problems (i.e., in terms of bounds), SNC is justifiably
proclaimed as a valuable alternative to the classical queueing
theory (see Ciucu and Schmitt~\cite{Ciucu12}). At the same time,
SNC is also justifiably questioned on the tightness of its bounds.
While the asymptotic tightness generally holds (see
Chang~\cite{Book-Chang}, p. 291, and Ciucu~\et~\cite{CiBuLi06}),
doubts on the bounds' numerical tightness shed skepticism on the
practical relevance of SNC. This skepticism is supported by the
fact that SNC largely employs the same probability methods as the
effective bandwidth theory, which was argued to produce largely
inaccurate results for non-Poisson arrival processes~(see
Choudhury~\et~\cite{choudhury96squeezing}). Moreover, although the
importance of accompanying bounds by simulations has already been
recognized in some early works (see~Zhang~\et~\cite{Zhang94} for
the analysis of GPS), the SNC literature is scarce in that
respect.

In this paper we reveal what is perhaps `feared' by SNC proponents
and expected by others: the bounds are very loose for the class of
MMOO processes, which is very relevant as these can be tuned for
various degrees of burstiness. In addition to providing numerical
evidence for this fact (the bounds can be off by arbitrary orders
of magnitude, e.g., by factors as large as $100$ or even $1000$),
we also prove that the bounds are asymptotically loose in
multiplexing regimes. Concretely, we (analytically) prove that the
bounds are `missing' an exponential decay factor of
${\mathcal{O}}\left(e^{-\alpha n}\right)$ in the number of flows
$n$, where $\alpha>0$; this missing factor was conjectured through
numerical experiments in Choudhury~\et~\cite{choudhury96squeezing}
in the context of effective bandwidth results (which scale
identically as the SNC bounds).

While this paper convincingly uncovers a major weakness in the SNC
literature, it also shows that the looseness of the bounds is
generally not inherent in SNC but it is due to the `temptatious'
but `poisonous' elementary tools from probability theory leveraged
in its application. We point out that such methods have also been
employed in the effective bandwidth literature dealing with
scheduling; see Courcoubetis and
Weber~\cite{courcoubetis95effective} for FIFO, Berger and
Whitt~\cite{Berger98} and Wischik~\cite{Wischik01} for SP,
Sivaraman and Chiussi~\cite{Chiussi99} for EDF, and
~Zhang~\et~\cite{Zhang94} and Bertsimas~\et~\cite{Bertsimas99} for
WFQ. Unlike the SNC results, which are given in terms of
non-asymptotic bounds, the corresponding effective bandwidth
results are typically given in larger buffer asymptotics regimes;
while exactly capturing the asymptotic decay rate, they fail to
capture the extra ${\mathcal{O}}\left(e^{-\alpha n}\right)$ decay
factor pointed out by Choudhury~\et~\cite{choudhury96squeezing} or
by Botvich and Duffield~\cite{Botvich95}.

To fix the weakness of existing SNC bounds, and also of existing
effective bandwidth asymptotic results in scheduling scenarios,
this paper leverages more advanced tools (i.e., martingale based
techniques) and derives new Martingale bounds improving
dramatically to the point of almost matching simulation results.
We show the improvements for per-flow delay bounds in FIFO, SP,
EDF, and WFQ scheduling scenarios with MMOO flows, and in addition
we prove the existence of the conjectured
${\mathcal{O}}\left(e^{-\alpha n}\right)$ decay factor. We point
out that extensions to more general Markovian arrivals are
immediate (see Appendix.A); due to their increased complexity,
however, the generalized results do not easily lend themselves to
visualizing the ${\mathcal{O}}\left(e^{-\alpha n}\right)$ decay
factor uncovered herein for MMOO flows.

The sharp bounds obtained in this paper are the first in the
conventional stochastic network calculus literature, i.e.,
involving service processes which decouple scheduling from the
analysis. Their significance, relative to existing sharp bounds in
the effective bandwidth literature (e.g.,
Duffield~\cite{Duffield94} and Chang~\cite{Book-Chang}, pp.
339-343, using martingale inequalities, or Liu~\et~\cite{Nain97}
by extending an approach of Kingman involving integral
inequalities~\cite{Kingman70}), is that they apply at the
\textit{per-flow} level for various scheduling; in turn, existing
sharp bounds only apply at the \textit{aggregate} level. In other
words, our sharp bounds generalize existing ones by accounting for
FIFO, SP, EDF, and WFQ scheduling.

The rest of the paper is structured as follows. In
Section~\ref{sec:pitfall} we identify, at an intuitive level, the
elementary tool from probability theory which is `responsible' for
the very loose (Standard) bounds in SNC. In
Section~\ref{sec:model} we describe the queueing model and some
necessary SNC formalisms. The core of the paper is
Section~\ref{sec:mmoo}, which computes the improved (Martingale)
and reviews the existing (Standard) SNC per-flow delay bounds in
multiplexing scenarios with MMOO flows; both analytical and
numerical comparisons of the bounds are further explored.
Concluding remarks are presented in Section~\ref{sec:conclusions}.

\section{Three Bounding Steps in SNC and One
Pitfall}\label{sec:pitfall} This section overviews the SNC
bounding approach to compute per-flow queueing metrics for broad
classes of arrivals and scheduling. In addition to identifying
three major steps in this approach, it is conveyed by means of a
simple example that careless bounding can lend itself to
impractical results.

Towards this end, we consider a simplified queueing systems in
which a (cumulative) arrival process $A(t)$ shares with some other
flows a server with capacity $C$ and infinite queue length. We are
particularly interested in the complementary distribution of
$A(t)$'s backlog process $B(t)$, which is bounded in SNC for some
$t,\sigma\geq0$ according to
\begin{equation}
\P\left(B(t)>\sigma\right)\leq\P\left(\sup_{0\leq s\leq
t}\left\{A(s,t)-S(s,t)\right\}>\sigma\right)~.\label{eq:genb}
\end{equation}
Here, $A(s,t):=A(t)-A(s)$ is the bivariate extension of $A(t)$,
whereas $S(s,t)$ is another bivariate process, called a
\textit{service process}, encoding the information about the
server, the scheduling, and the other arrival processes that
$A(t)$ shares the server with. In the simplest setting with no
other arrivals, $S(s,t)=C(t-s)$ and Eq.~(\ref{eq:genb}) (with
equality) recovers Reich's equation. In another setting in which
$A(t)$ receives the lowest priority, should the server implement a
static priority (SP) scheduler, then $S(s,t)=C(t-s)-A_c(s,t)$,
where $A_c(s,t)$ denotes the other arrivals at the server.

Eq.~(\ref{eq:genb}) typically continues in SNC by invoking the
Union Bound, i.e.,
\begin{equation}
\textrm{Eq.}~(\ref{eq:genb})\ldots\leq\sum_{s=0}^t\P\left(A(s,t)-S(s,t)>\sigma\right)~.\label{eq:genboole}
\end{equation}
The probability events can be further computed either by 1)
convolving the distribution functions of $A(s,t)$ and $S(s,t)$,
when available, and under appropriate independence assumptions, or
by following a more elegant procedure using the Chernoff bound,
i.e.,
\begin{equation}
\textrm{Eq.}~(\ref{eq:genboole})\ldots\leq\sum_{s=0}^tE\left[e^{\theta\left(A(s,t)-S(s,t)\right)}\right]e^{-\theta\sigma}~,\label{eq:genchernoff}
\end{equation}
for some $\theta>0$. The expectation can be split into a product
of expectations, according to the statistical independence
properties of $A(s,t)$ and $S(s,t)$, and the sum can be further
reduced to some canonical form.

Eqs.~(\ref{eq:genb})-(\ref{eq:genchernoff}) outline three major
bounding steps. The first is `proprietary' to SNC, in the sense
that it involves the unique construction of a `proprietary'
service process $S(s,t)$ which decouples scheduling from analysis.
The next two follow general purpose methods in probability theory,
which are applied in the same form in the effective bandwidth
theory, except that $S(s,t)$ is now a random process rather than a
constant-rate function.

The second step in particular reveals a convenient mathematical
continuation of Eq.~(\ref{eq:genb}). The reason for this
`temptatious' step to be consistently invoked in SNC stems from
the `freedom' of seeking for bounds rather than exact results. As
we will show over the rest of this section, and of the paper, this
`temptatious' step is also `poisonous' in the sense that it lends
itself to very loose bounds for a class of queueing scenarios
which is being proclaimed in SNC as a highlight of its scope.

To convey insight into this direction, let us consider the
stationary but non-ergodic process
\begin{equation}
A(s,t)=(t-s)X~\forall 0\leq s\leq t~,\label{eq:advcons}
\end{equation}
where $X$ is a Bernoulli random variable taking values in
$\{0,2\}$, each with probabilities $1-\eps>.5$ and $\eps>0$.
Assume also that $S(s,t)=t-s$. Clearly, for $\sigma>0$ and for
sufficiently large $t$, the backlog process satisfies
\begin{equation*}
\P\left(B(t)>\sigma\right)=\eps~.
\end{equation*}
In turn, the application of the bound from Eq.~(\ref{eq:genboole})
lends itself to a bogus bound, i.e.,
\begin{equation*}
\P\left(B(t)>\sigma\right)\leq \eps t~,
\end{equation*}
for $\sigma<1$ (for $\sigma\geq 1$, the bound diverges as well).
The underlying reason behind this bogus result is that the Union
Bound from Eq.~(\ref{eq:genboole}) is agnostic to the statistical
poperties of the increments of the arrival process $A(s,t)$.

The construction of $A(s,t)$ from Eq.~(\ref{eq:advcons}) is meant
to convey insight into the poor performance of the Union Bound for
arrivals with correlated increments, such as MMOO processes.
Within the same class, another relevant arrival process is the
fractional Brownian motion which has long-range correlations; the
analysis of such process was done either by approximations (e.g.,
Norros~\cite{Nor95}) or by using the Union Bound (e.g., Rizk and
Fidler~\cite{Rizk12_FBM}). The rest of the paper will
unequivocally reveal that the Union Bound leads to very loose
per-flow bounds for MMOO processes.

The Union Bound can however lend itself to reasonably tight bounds
when $X_s:=A(s,t)$'s are rather uncorrelated (see
Talagrand~\cite{tala96b}). Shroff and Schwartz~\cite{Shroff98b}
argued that the effective bandwidth theory yields reasonable
bounds only for Poisson processes. Moreover, Ciucu~\cite{Ciucu07}
provided numerical evidence that SNC lends itself to reasonably
tight bounds for Poisson arrivals as well.

\section{Queueing Model}\label{sec:model}
This section introduces the queueing model and necessary SNC
formalisms. The time model is continuous. Consider a stationary
(bivariate) arrival process $A(s,t)$ defined as
\begin{equation*}
A(s,t):=\int_{u=s}^t a(u)du~\forall 0\leq s\leq t~,~~~
A(t):=A(0,t)~,
\end{equation*}
where $a(s)~\forall s\geq0$ is the increment process.

According to Kolmogorov's extension theorem, the one-side
(stationary) process $\left\{a(s):0\leq s<\infty\right\}$ can be
extended to a two-side process
$\left\{a(s):-\infty<s<\infty\right\}$ with the same distribution.
For convenience, we often work with the \textit{reversed}
cumulative arrival process $\Ar(s,t)$ defined as
\begin{equation*}
\Ar(s,t)=\int_{u=s}^ta(-u)du~\forall 0\leq s\leq t~,~~~
\Ar(t):=\Ar(0,t)~.
\end{equation*}
This definition is identical with that of $A(s,t)$, except that
the time direction is reversed.

Working with time reversed processes is particularly convenient in
that the steady-state queueing process (say in a queueing system
with constant-rate capacity $C$ fed by the one-side increment
process $a(s)$) can be represented by Reich's equation
\begin{equation*}
Q=\sup_{t\geq0}\left\{\Ar(t)-Ct\right\}~.
\end{equation*}
The evaluation of $Q$ needs an additional \textit{stability}
condition, e.g.,
$\limsup_{t\rightarrow\infty}\frac{A^{\textrm{r}}(t)}{t}<C~\textrm{a.s.}$
(see Chang~\cite{Book-Chang}, pp. 293-294); this condition is
fulfilled by the (stronger) Loynes' condition, i.e., $a(s)$ is
also ergodic and
$\lim_{t\rightarrow\infty}\frac{A(t)}{t}=E\left[a(1)\right]<C~\textrm{a.s.}$

\begin{figure}[t]
\centering
 \includegraphics[scale=0.5]{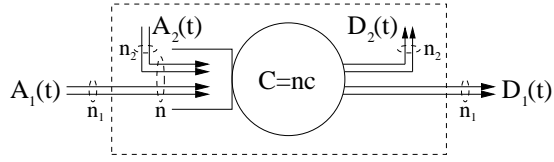}
\caption{A queueing system with two arrival processes $A_1(t)$ and
$A_2(t)$, each containing $n_1$ and $n_2$ sub-flows. The server
has a capacity $C=nc$, where $n=n_1+n_2$. We are interested in the
delay distribution of $A_1(t)$. }\label{fig:qsystem}
\end{figure}

In this paper we mostly consider the queueing system depicted in
Figure~\ref{fig:qsystem}. Two cumulative arrival processes
$A_1(t)$ and $A_2(t)$, each containing $n_1$ and $n_2$ sub-flows,
are served by a server with constant-rate $C=nc$, where
$n=n_1+n_2$. The parameter $c$ is referred to as the
\textit{per-(sub)flow capacity}, and will be needed in the context
of asymptotic analysis. For clarity, $A_1(t)$ and $A_2(t)$ will
also be suggestively referred to as the \textit{through} and
\textit{cross} (aggregate) flows, respectively. The data units are
infinitesimally small and are referred to as \textit{bits}. The
queue has an infinite size capacity, and is assumed to be stable.

The performance measure of interest is the \textit{virtual} delay
process for the (through) flow $A_1(t)$, defined as
\begin{equation*}
W_1(t):=\inf\left\{d\geq0:A_1(t-d)\leq D_1(t)\right\}~\forall
t\geq0~,
\end{equation*}
where $D_1(t)$ is the corresponding departure process of $A_1(t)$
(see Figure~\ref{fig:qsystem}). The attribute \textit{virtual}
expresses the fact that $W_1(t)$ models the delay experienced by a
\textit{virtual} data unit departing at time $t$. Note that
$W_1(t)$ is the horizontal distance between the curves $A_1(t)$
and $D_1(t)$, starting backwards from the point $(t,D_1(t))$ in
the Euclidean space.

In stochastic network calculus, queueing performance metrics
(e.g., bounds on the distribution of the delay process $W_1(t)$)
are derived by constructing \textit{service curve} processes,
which relate the departure and arrival processes by a
$(\textrm{min},+)$ convolution. For instance, in the case of
$A_1(t)$ and $D_1(t)$, the corresponding service process is a
stochastic process $S_1(s,t)$ such that
\begin{equation}
D_1(t)\geq A_1\conv S_1(t)~\forall t\geq0~,\label{eq:sc}
\end{equation}
where `$\conv$' is the $(\textrm{min},+)$ convolution operator
defined for all sample-paths as $A_1\conv S_1(t):=\inf_{0\leq
s\leq t}\left\{A_1(s)+S_1(s,t)\right\}$.

The service process $S_1(s,t)$ typically encodes the information
about the cross aggregate $A_2(t)$ and the scheduling algorithm;
other information such as the packet size distribution is omitted
herein in accordance to the infinitesimal data units assumption.
Conceptually, the service process representation from
Eq.~(\ref{eq:sc}) encodes $A_1(t)$'s own service view, as if it
was alone at the network node (i.e., not competing for the service
capacity $C$ with other flows). Although the representation is not
exact due to the inequality from Eq.~(\ref{eq:sc}), it suffices
for the purpose of deriving upper bounds on the distribution of
$W_1(t)$. The driving key property is that Eq.~(\ref{eq:sc}) holds
for \textit{all} arrival processes $A_1(t)$. Due to this property,
the service representation in SNC is \textit{somewhat analogous}
with the impulse-response representation of signals in linear and
time invariant (LTI) systems (see Ciucu and Schmitt~\cite{Ciucu12}
for a recent discussion on this analogy).

In this paper we will compute the distribution of the through
aggregate's delay process $W_1(t)$ for four distinct scheduling
algorithms at the server, i.e., First-In-First-Out (FIFO), Static
Priority (SP), Earliest-Deadline-First (EDF), and Generalized
Processor Sharing (GPS). The enabling service process $S_1(s,t)$
for the delay computations, for each of scheduling algorithms,
will be presented in Section~\ref{sec:sncmart}.

\section{SNC Bounds for MMOO Processes}\label{sec:mmoo}
In this section we consider the queueing scenario from
Figure~\ref{fig:qsystem}, in which the sub-flows comprising
$A_1(t)$ and $A_2(t)$ are Markov-Modulated On-Off (MMOO)
processes. Because such processes can be tuned for various degrees
of burstiness, they are particularly relevant for both modelling
purposes and testing the tightness of related performance bounds.

After defining the MMOO processes, we derive Martingale bounds for
the distribution of $W_1(t)$ for FIFO, SP, EDF, and GPS
scheduling. Then we overview the corresponding Standard bounds in
SNC. Lastly, we compare these bounds both asymptotically, as well
as against simulations.

\begin{figure}[h]
\centering
 \includegraphics[scale=0.75]{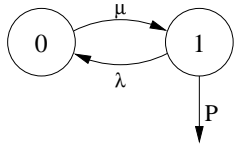}
\caption{A Markov-modulated On-Off (MMOO) process
}\label{fig:onoff}
\end{figure}

Each MMOO sub-flow is modulated by a continuous time Markov
process $Z(t)$ with two states denoted by $0$ and $1$, and
transition rates $\mu$ and $\lambda$ as depicted in
Figure~\ref{fig:onoff}. The cumulative arrival process for each
sub-flow is defined as
\begin{equation}
A'(s,t):=\int_{u=s}^tZ(u)Pdu~\forall 0\leq s\leq
t~,~A'(t):=A'(0,t)~,\label{eq:mmoosf}
\end{equation}
where $P>0$ is the peak rate. In other words, $A'(t)$ models a
data source transmitting with rates $0$ and $P$ while $Z(t)$
delves in the $0$ and $1$ states, respectively. The steady-state
`On' probability is $p:=\frac{\mu}{\lambda+\mu}$ and the average
rate is $pP$.

\begin{figure}[h]
\centering
 \includegraphics[scale=0.6]{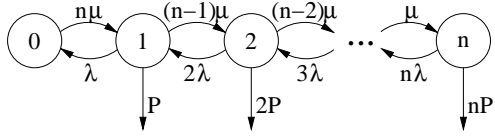}
\caption{A Markov-modulated process for the aggregation of $n$
homogeneous MMOOs }\label{fig:nonoff}
\end{figure}

When $n$ such statistically independent sources are multiplexed
together then the corresponding modulating Markov process, denoted
with abuse of notation as $Z(t)$ as well, has the states
$\left\{0,1,\dots,n\right\}$ and the transition rates as depicted
in Figure~\ref{fig:nonoff}. The cumulative arrival process for the
aggregate flow is defined identically as for each sub-flow, i.e.,
\begin{equation*}
A(s,t):=\int_{u=s}^tZ(u)Pdu~\forall 0\leq s\leq t~,~A(t):=A(0,t)~.
\end{equation*}
Note that, by definition, $A(s,t)$ is continuous.

\subsection{Martingale Bounds}\label{sec:sncmart}
Recall our main goal of deriving bounds on the distribution of the
through aggregate's delay process $W_1(t)$ for the FIFO, SP, EDF,
and GPS scheduling scenarios in the network model from
Figure~\ref{fig:qsystem}. We start this section with a general
result enabling the analysis of all four scheduling scenarios, and
then analyze each separately.

\begin{theorem}{({\sc{Martingale Sample-Path Bound}})}\label{th:tspb}
Consider the single-node queueing scenario from
Figure~\ref{fig:qsystem}, in which $n$ sub-flows are statistically
independent MMOO processes with transition rates $\mu$ and
$\lambda$, and peak rate $P$, and starting in the steady-state.
The aggregate arrival processes are $A_1(t)$ and $A_2(t)$, each
being modulated by the (stationary) Markov processes $Z_1(t)$ and
$Z_2(t)$ with $n_1$ and $n_2$ states, respectively, with
$n_1+n_2=n$. Assume that the utilization factor
$\rho:=\frac{pP}{c}$ satisfies $\rho<1$ for stability, where $p$
is the steady-state `On' probability; assume also that $P>c$ to
avoid a trivial scenario with zero delay. Then the following
sample-path bound holds for all $0\leq u\leq t$ and $\sigma$
\begin{eqnarray}
&&\hspace{-1cm}\P\left(\sup_{0\leq s<
t-u}\left\{A_1(s,t-u)+A_2(s,t)-C(t-s)\right\}>\sigma\right)\notag\\
&&\hspace{1cm}\leq K^ne^{-\gamma\left(C_1
u+\sigma\right)}~,\label{eq:tspb}
\end{eqnarray}
where $C_1=n_1c$,
$K=\rho\left(\frac{\rho-p}{1-p}\right)^{\frac{p}{\rho}-1}$, and
$\gamma=\frac{(\lambda+\mu)(1-\rho)}{P-c}$.
\end{theorem}

\medskip

We point out that the crucial element in the sample-path bound
from Eq.~(\ref{eq:tspb}) is the \textit{parameter} $u$, which can
be explicitly \textit{tuned} depending on the scheduling algorithm
for the bits of $A_1(t)$ and $A_2(t)$. From a conceptual point of
view, the parameter $u$ encodes the information about the
underlying scheduling, whereas the theorem further enables the
\textit{per-flow} delay analysis for several common scheduling
algorithms: FIFO, SP, and EDF (see
Subsections~\ref{sec:fifo}--\ref{sec:edf}).

The obtained delay bounds generalize the delay bounds previously
obtained by Palmowski and Rolski~\cite{Palmowski96}, by further
accounting for several scheduling algorithms\footnote{More
exactly, \cite{Palmowski96} gives backlog bounds at the aggregate
level which can be immediately translated into delay bounds, given
the fixed server capacity for the whole aggregate.}. The bounds
from~\cite{Palmowski96} can be recovered by applying
Theorem~\ref{th:tspb} with $A_2(t)=0$ (i.e., no cross traffic and
thus no scheduling being considered). The key to the proof of
Theorem~\ref{th:tspb} is the construction of a single martingale
$M_t$ from two others suitably shifted in time; this subtle
construction, together with the scheduling abstraction feature of
SNC, are instrumental to the per-flow analysis for the different
scheduling algorithms. Furthermore, the sample-path bound from
Eq.~(\ref{eq:tspb}) follows from a standard technique based on the
Optional Sampling theorem, applied to the martingale $M_t$; for
relevant definitions and results related to martingales we refer
to the Appendix.B. Also, for the generalization of
Theorem~\ref{th:tspb} to general Markov fluid processes we refer
to Appendix.A; as mentioned in the Introduction, however, the
generalized result does not lend itself to visualizing the
conjectured ${\mathcal{O}}\left(e^{-\alpha n}\right)$ decay
factor, for which reason we mainly focus on MMOO processes.

\proof Fix $u\geq0$ and $\sigma$. For convenience, let us bound
the probability from Eq.~(\ref{eq:tspb}) by shifting the time
origin and using the time-reversed representation of arrival
processes described in Section~\ref{sec:model}, i.e.,
\begin{eqnarray}
\P\left(\sup_{t>u}\left\{\Ar_1(u,t)+\Ar_2(u,t)-C(t-u)\right\}+\Ar_2(u)-C_2u>C_1u+\sigma\right)~,\label{eq:revspb}
\end{eqnarray}
where $C_2=n_2c$. This representation is possible because the
underlying Markov modulating processes of $A_1(t)$ and $A_2(t)$,
i.e., $Z_1(t)$ and $Z_2(t)$, respectively, are
\textit{time-reversible} processes (see, e.g.,
Mandjes~\cite{Mandjes07}, p. 57); the reversibility is a
consequence of the fact that $Z_1(t)$ and $Z_2(t)$ are stationary
birth-death processes (see Kelly~\cite{Kelly79}, pp. 10-11).
Denote by $\Zr_1(t)$ and $\Zr_2(t)$ the time-reversed versions of
$Z_1(t)$ and $Z_2(t)$, respectively.

Given the previous probability event we define the stopping time
\begin{eqnarray}
T:=\inf\left\{t>u:\Ar_1(u,t)+\Ar_2(u,t)-C(t-u)+\Ar_2(u)-C_2u>C_1u+\sigma\right\}~.\label{eq:stot}
\end{eqnarray}
This construction is motivated by the fact that
$\P\left(T<\infty\right)$ is exactly the probability from
Eq.~(\ref{eq:revspb}). The goal of the rest of the proof is to
bound $\P\left(T<\infty\right)$.

Let $\P_{i,j}$ denote the underlying probability measure
conditioned on $\Zr_1(u)=i$ and $\Zr_2(0)=j$, for $0\leq i\leq
n_1$ and $0\leq j\leq n_2$. Denote also the stationary probability
vectors of $\Zr_1(u)$ and $\Zr_2(u)$ by
$(\pi_{1,0},\dots,\pi_{1,n_1})$ and
$(\pi_{2,0},\dots,\pi_{2,n_2})$, respectively.

Next we define the following two processes
\begin{eqnarray*}
M_1(t)&:=&e^{-\theta\left(\Zr_1(t)-i\right)}e^{\gamma\int_u^t\left(P\Zr_1(s)-C_1\right)ds}~\forall
t\geq u~\textrm{and}\\
M_2(t)&:=&e^{-\theta\left(\Zr_2(t)-j\right)}e^{\gamma\int_0^t\left(P\Zr_2(s)-C_2\right)ds}~\forall
t\geq0~,
\end{eqnarray*}
where $\theta:=\log{\frac{\mu}{\lambda}\frac{P-c}{c}}$. Note that
$\theta<0$ due to the stability condition $\rho<1$.

According to Palmowski and Rolski~\cite{Palmowski96}, both
$M_1(t)$ and $M_2(t)$ are martingales with respect to (wrt)
$\P_{i,j}$ and the natural filtration (for the original result see
Ethier and Kurtz~\cite{EK86}, p. 175). Moreover, according to
Lemmas~\ref{lm:pim} and~\ref{lm:os} from the Appendix, the
following process
\begin{equation*}
M_t:=\left\{\begin{array}{lcc}M_2(t)&,&t\leq u\\M_1(t)M_2(t)&,&t>
u\end{array}\right.
\end{equation*}
is also a martingale (note that $M_1(u)=1$, by construction).

Because $T$ may be unbounded, we need to construct the bounded
stopping times $T\wedge k$ for all $k\in\N$. For these times, the
Optional Sampling theorem (see Theorem~\ref{th:ost} in the
Appendix) yields
\begin{eqnarray*}
E_{i,j}\left[M_0\right]=E_{i,j}\left[M_{T\wedge k}\right]~,
\end{eqnarray*}
for all $k\in N$, where the expectations are taken wrt $\P_{i,j}$.
Using $E_{i,j}\left[M_0\right]=1$ and according to the
construction of $M_2(t)$ we further obtain for $k>u$
\begin{eqnarray*}
1&\geq&E_{i,j}\left[M_{T\wedge k}I_{\left\{T\leq k\right\}}\right]\\
&\geq&
e^{-\theta\left(\frac{C_1+C_2}{P}-(i+j)\right)}e^{\gamma\left( C_1
u+\sigma\right)}\P_{i,j}\left(T\leq k\right)~,
\end{eqnarray*}
where $I_{\left\{\cdot\right\}}$ denotes the indicator function.
The first term in the product follows from $\theta<0$ and
\begin{equation*}\left(\Zr_1(T)+\Zr_2(T)\right)P\geq C_1+C_2~,
\end{equation*}according to the construction of $T$ from Eq.~(\ref{eq:stot})
and the continuity property of the arrival processes. The second
term follows from $\gamma>0$ and the construction of $T$.

By deconditioning on $i$ and $j$ (note that $\Zr_1(u)$ and
$\Zr_2(0)$ are in steady-state by construction) we obtain
\begin{equation*}
\P\left(T\leq
k\right)\leq\sum_{i,j}\pi_{1,i}\pi_{2,j}e^{\theta\left(\frac{C_1+C_2}{P}-(i+j)\right)}e^{-\gamma\left(
C_1 u+\sigma\right)}~.
\end{equation*}
Using the identities
\begin{equation*}\sum_{i=0}^{n_1}\pi_{1,i}e^{\theta\left(\frac{C_1}{P}-i\right)}=K^{n_1}~\textrm{and}~\sum_{j=0}^{n_2}\pi_{2,j}e^{\theta\left(\frac{C_2}{P}-j\right)}=K^{n_2}
\end{equation*}
(see~\cite{Palmowski96}) and taking $k\rightarrow\infty$ we
finally obtain that
\begin{equation*}
\P\left(T<\infty\right)\leq K^{n}e^{-\gamma\left(C_1
u+\sigma\right)}~,
\end{equation*}
which completes the proof.~\hfill $\Box$

\medskip

In the following we fix $0\leq d\leq t$ and derive bounds on
$\P\left(W_1(t)>d\right)$ for FIFO, SP, EDF, and GPS scheduling;
the derivations follow more or less directly by instantiating the
parameters of Theorem~\ref{th:tspb} for each scheduling case.

\subsubsection{FIFO}\label{sec:fifo}
The FIFO server schedules the data units of $A_1(t)$ and $A_2(t)$
in the order of their arrival times.

To derive a bound on the distribution of the through aggregate's
(virtual) delay process $W_1(t)$, we rely on a service process
construction for FIFO scheduling, as mentioned in
Section~\ref{sec:model}. We use the service process from
Cruz~\cite{Cruz98} extended to bivariate stochastic processes,
i.e.,
\begin{equation}
S_1(s,t)=\left[C(t-s)-A_2(s,t-x)\right]_+I_{\left\{t-s>x\right\}}~,\label{eq:fifosc}
\end{equation}
for some fixed $x\geq0$ and independent of $s$ and $t$ (for a
proof, in the sightly simpler case of univariate processes, see~Le
Boudec and Thiran~\cite{Book-LeBoudec}, pp. 177-178; for a more
recent and general proof see Liebeherr~\et~\cite{Yashar11}). By
notation, $[y]_+:=\max\left\{y,0\right\}$ for some real number
$y$.

Using the equivalence of events
\begin{equation*}
W_1(t)>d\Leftrightarrow A_1(t-d)>D_1(t)~,
\end{equation*}
and also the service process representation from
Eq.~(\ref{eq:sc}), we can bound the distribution of $W_1(t)$ as
follows
\begin{eqnarray}
&&\hspace{-0.5cm}\P\left(W_1(t)>d\right)\notag\\
&&\leq\P\left(A_1(t-d)>A_1\conv
S_1(t)\right)\notag\\
&&=\P\Big(\sup_{0\leq
s<t-d}\left\{A_1(s,t-d)-\left[C(t-s)-A_2(s,t-x)\right]_+I_{\left\{t-s>x\right\}}\right\}>0\Big\}~.\label{eq:derivfifo}
\end{eqnarray}
Here we restricted the range of $s$ from $[0,t]$ to $[0,t-d)$,
using the positivity of the `$[\cdot]_+$' operator and the
monotonicity of $A_1(s,t)$.

Because $x$ is a free parameter in the FIFO service process
construction from Eq.~(\ref{eq:fifosc}), let us choose $x=d$. With
this choice it follows from above that
\begin{eqnarray*}
&&\hspace{-0.5cm}\P\left(W_1(t)>d\right)\\
&&\leq\P\left(\sup_{0\leq
s<t-d}\left\{A_1(s,t-d)+A_2(s,t-d)-C(t-s)\right\}>0\right)~.
\end{eqnarray*}
By applying Theorem~\ref{th:tspb} with $u=0$ and $\sigma=Cd$, we
get the following\newline \medskip \textbf{Martingale Delay Bound
(FIFO)}:
\begin{equation}
\P\Big(W_1(t)>d\Big)\leq K^{n}e^{-\gamma Cd}~,\label{eq:fifodb}
\end{equation}
where $K$ and $\gamma$ are given in Theorem~\ref{th:tspb}. Note
that the bound is invariant to the number of sub-flows $n_1$,
which is a property characteristic to a \textit{virtual} delay
process (for FIFO); such a dependence will be established by
changing the measure from a virtual delay process to a packet
delay process (see Section~\ref{sec:simus}).

\subsubsection{SP}\label{sec:SP}
Here we consider an SP server giving higher priority to the data
units of the cross flow $A_2(t)$. We are further interested in the
delay distribution of the lower priority flow; the case of the
higher priority flow is a consequence of the previous FIFO result.

We follow the same procedure of first encoding $A_1(t)$'s service
view in a service process, e.g., (see Fidler~\cite{Fidler06}),
\begin{equation}
S_1(s,t)=C(t-s)-A_2(s,t)~.\label{eq:sprocsp}
\end{equation}
now in the case of SP scheduling.

To bound the distribution of $W_1(t)$ we continue the first two
lines of Eq.~(\ref{eq:derivfifo}) as follows
\begin{eqnarray*}
&&\hspace{-0.5cm}\P\left(W_1(t)>d\right)\\
&&\leq\P\left(\sup_{0\leq
s<t-d}\left\{A_1(s,t-d)+A_2(s,t)-C(t-s)\right\}>0\right)~.
\end{eqnarray*}
By applying Theorem~\ref{th:tspb} with $u=d$ and $\sigma=0$, we
get the following\newline \textbf{Martingale Delay Bound (SP)}:
\begin{equation}
\P\Big(W_1(t)>d\Big)\leq K^{n}e^{-\gamma C_1d}~,\label{eq:spdb}
\end{equation}
where $K$ and $\gamma$ are given in Theorem~\ref{th:tspb}. Note
that, as expected, the SP delay bound recovers the FIFO delay
bound from Eq.~(\ref{eq:fifodb}) when there is no cross aggregate,
i.e., in the case when $C_1=C$.

\subsubsection{EDF}\label{sec:edf}
An EDF server associates the relative deadlines $d_1^*$ and
$d_2^*$ with the data units of $A_1(t)$ and $A_2(t)$,
respectively. Furthermore, all data units are served in the order
of their remaining deadlines, even when they are negative (we do
not consider data unit losses).

A service process for $A_1(t)$ is for some $x>0$
\begin{equation}
S_1(s,t)=\left[C(t-s)-A_2(s,t-x+\min\{x,y\})\right]_+I_{\left\{t-s>x\right\}}~,\label{eq:edfsc}
\end{equation}
where $y=d_1^*-d_2^*$ (see Liebeherr~\et~\cite{Yashar11}). This
service process generalizes the FIFO one from
Eq.~(\ref{eq:fifosc}) (which holds for $y=0$, i.e., the associated
deadlines to the flows are equal), and it also generalizes a
previous EDF service process by Li~\et~\cite{LiBuLi07} (which is
restricted to $x=0$).

To derive a bound on $\P\left(W_1(t)>d\right)$, for some $d\geq0$,
let us first choose $x:=d$, as we did for FIFO. Next we
distinguish two cases depending on the sign of $y$.

If $y\geq0$ then the continuation of Eq.~(\ref{eq:derivfifo}) is
\begin{eqnarray*}
&&\hspace{-0.5cm}\P\left(W_1(t)>d\right)\\
&&\leq\P\left(\sup_{0\leq
s<t-d}\big\{A_1(s,t-d)+A_2(s,t-d+\min\{d,y\})-C(t-s)\right\}>0\Big)~.
\end{eqnarray*}
By changing the variable $t\leftarrow t+d-\min\{d,y\}$ we get
\begin{eqnarray*}
&&\hspace{-0.5cm}\P\left(W_1(t)>d\right)\\
&&\leq\P\left(\sup_{0\leq
s<t-\min\{d,y\}}\left\{A_1(s,t-\min\{d,y\})+A_2(s,t)-C(t-s+d-\min\{d,y\})\right\}>0\right)~.
\end{eqnarray*}
(we point out that as we are looking for the steady-state
distribution of $W_1(t)$, we can omit the technical details of
writing $W_1(t+d-\min\{d,y\})$ above.) We can now apply
Theorem~\ref{th:tspb} with $u=\min\{d,y\}$ (note that both $d$ and
$y$ are positive) and $\sigma=C(d-\min\{d,y\})$, and get the
following\newline \textbf{Martingale Delay Bound (EDF) ($d_1^*\geq
d_2^*$ Case)}:
\begin{equation}
\P\Big(W_1(t)>d\Big)\leq K^{n}e^{\gamma
C_2\min\left\{d_1^*-d_2^*,d\right\}}e^{-\gamma
Cd}~,\label{eq:edfdb1}
\end{equation}
where $K$ and $\gamma$ are given in Theorem~\ref{th:tspb}.

The second case, i.e., $y<0$, is slightly more complicated. The
reason is that $\min\{d,y\}=y$ (see Eq.~(\ref{eq:edfsc})) such
that the continuation of Eq.~(\ref{eq:derivfifo}) becomes
\begin{eqnarray}
&&\hspace{-1cm}\P\left(\sup_{0\leq
s<t-d}\left\{A_1(s,t-d)-\left[C(t-s)-A_2(s,t-d+y)\right]_+I_{\left\{t-s>d\right\}}\right\}>0\right)~.\label{eq:derivedf}
\end{eqnarray}
Note that when $s\in\left[t-d+y,t-d\right)$, then one must
consider $A_2(s,t-d+y):=0$ according to the conventions
from~\cite{Yashar11}. Therefore, one must perform the splitting
$[0,t-d)=[0,t-d+y)\cup[t-d+y,t-d)$; thereafter, by changing the
variable $t\leftarrow t+d$, the continuation of
Eq.~(\ref{eq:derivedf}) is
\begin{eqnarray*}
&&\leq\P\left(\max\left\{\sup_{0\leq
s<t+y}\left\{A_2(s,t+y)+A_1(s,t)-C(t-s)\right\},\sup_{t+y\leq
s<t}\left\{A_1(s,t)-C(t-s)\right\}\right\}>Cd\right)\\
&&\leq\P\left(\sup_{0\leq
s<t+y}\left\{A_2(s,t+y)+A_1(s,t)-C(t-s)\right\}>Cd\right)+\P\left(\sup_{0\leq
s<t}\left\{A_1(s,t)-C(t-s)\right\}>Cd\right)
\end{eqnarray*}
In the third line we applied the Union Bound [\textit{sic}], which
is conceivably tight because the two elements in the `$\max$' are
rather uncorrelated. Moreover, we extended the left margin in the
last supremum (in the fourth line), as we are looking for upper
bounds, whereas the martingale argument from Theorem~\ref{th:tspb}
is insensitive to where the left margin starts.

The last two probabilities can be directly evaluated with
Theorem~\ref{th:tspb}. For the first one we set $u=-y$ (note that
$y$ is now negative) and $\sigma=Cd$. For the second one we set
$u=0$, $n_2=0$, $\sigma=Cd$, and we properly rescale the per-flow
capacity $c$ and utilization factor $\rho$ (see below). In this
way get the following\newline \textbf{Martingale Delay Bound (EDF)
($d_1^*< d_2^*$ Case)}:
\begin{equation}
\P\Big(W_1(t)>d\Big)\leq K^{n}e^{\gamma
C_2\left(d_1^*-d_2^*\right)}e^{-\gamma Cd}+K'^{n}e^{-\gamma'
Cd}~,\label{eq:edfdb2}
\end{equation}
with the same $K$ and $\gamma$ from Theorem~\ref{th:tspb}, whereas
$K'$ and $\gamma'$ are obtained alike $K$ and $\gamma$, but after
rescaling $c'\leftarrow\frac{n_1+n_2}{n_1}c$ and
$\rho'=\frac{n_1}{n_1+n_2}\rho$.

Note that the first EDF bound from Eq.~(\ref{eq:edfdb1}) recovers
the FIFO bound when the associated deadlines are equal, i.e., when
$d_1^*=d_2^*$. In turn, the second EDF bound from
Eq.~(\ref{eq:edfdb1}) would also recover the FIFO bound, but only
by dispensing with the unnecessary splitting of the interval
$[0,t-d)$ since $y=0$.

\subsubsection{GPS}\label{sec:gps}
We consider a GPS server (see Parekh and Gallager~\cite{Parekh93},
or Demers~\et~\cite{Demers89}) which assigns positive weights
$\phi_1$ and $\phi_2$, normalized here such that
$\phi_1+\phi_2=1$, to the flows $A_1(t)$ and $A_2(t)$,
respectively. Denoting by $D_1(s,t)$ and $D_2(s,t)$ the
corresponding departure processes, GPS guarantees that for every
continuous backlogged flow $i$ in a time interval $[s,t)$, the
following holds
\begin{equation}
\frac{D_i(s,t)}{D_j(s,t)}\geq\frac{\phi_i}{\phi_j}~\textrm{for}~j\neq
i~.\label{eq:gpsprop}
\end{equation}

One service process for the flow $A_1(t)$ is
(see~Chang~\cite{Book-Chang}, p. 68)
\begin{equation}
S_1(s,t)=\phi_1 C(t-s)~,\label{eq:sgps}
\end{equation}
which corresponds to the minimum service guarantee by the GPS
property from Eq.~(\ref{eq:gpsprop}). Unfortunately, this service
process does not capture the full server capacity allocated to the
flow $A_1(t)$ when the other flow is not backlogged, and it thus
conceivably leads to loose bounds.

An improved service process was constructed by
Li~\et~\cite{LiBuLi07}, but it requires an additional concavity
assumption on the flow $A_2(t)$; therefore, it does not apply in
our setting. An improvement in the general case (i.e., for any
types of arrivals) can be obtained when the SP service process
from Eq.~(\ref{eq:sprocsp}) is larger than the one from
Eq.~(\ref{eq:sgps}); note that the SP service process holds by
default. The improvement can be substantial for small values of
$\phi_1$. Indeed, in the extreme case when $\phi_1=0$, the GPS
service process from Eq.~(\ref{eq:sgps}) would predict infinite
delays as the system would be (wrongly) perceived in overload; in
turn, by relying on the SP service process, finite delays can be
obtained.

Nevertheless, despite the pessimistic outlook of relying on the
service process from Eq.~(\ref{eq:sgps}), we will compute the
Martingale and Standard bounds on the distribution of $W_1(t)$.
The reason is to illustrate, by means of comparisons with
simulations, that the service process from Eq.~(\ref{eq:sgps})
lends itself to loose bounds, even in `favorable' rate
proportional processor sharing (RPPS) scenarios such as $n_1=n_2$
and $\phi_1=\phi_2$ (see Section~\ref{sec:simus}).

To derive the Martingale delay bound, we continue the first two
lines of Eq.~(\ref{eq:derivfifo}) as follows
\begin{eqnarray*}
&&\hspace{-0.9cm}\P\left(W_1(t)>d\right)\\
&&\hspace{-0.5cm}\leq\P\left(\sup_{0\leq
s<t-d}\left\{A_1(s,t-d)-\phi_1C(t-s)\right\}>0\right)~.
\end{eqnarray*}
Next, by applying Theorem~\ref{th:tspb} with $C:=\phi_1C$, $u:=d$,
$\sigma:=\phi_1Cd$, and $n_2:=0$, in this order, we obtain the
following\newline \textbf{Martingale Delay Bound (GPS)}:
\begin{equation}
\P\Big(W_1(t)>d\Big)\leq K^{n}e^{-\gamma
\phi_1Cd}~,\label{eq:gpsdb}
\end{equation}
where $K$ and $\gamma$ are given in Theorem~\ref{th:tspb}; note
that the new utilization factor is $\rho:=\frac{n_1pP}{\phi_1C}$.

\subsection{Standard Bounds}
Here we briefly review the standard (per-flow) delay bounds
obtained with SNC for FIFO, SP, EDF, and GPS scheduling
algorithms. These bounds will be compared, both analytically and
numerically, against the Martingale bounds computed so far.

We assume that for each MMOO sub-flow $A_0(t)$ the corresponding
Markov process $Z(t)$ (with two states, from
Figure~\ref{fig:onoff}) starts in the steady-state, i.e.,
\begin{equation*}
\P\left(Z(0)=0\right)=1-p~\textrm{and}~\P(Z(0)=1)=p~,
\end{equation*}
where $p$ was defined earlier, i.e., $p=\frac{\mu}{\lambda+\mu}$.
The computation of the Standard delay bounds relies on the moment
generating function (MGF) of $A_0(t)$, which can be written for
all $t\geq0$ and some $\theta>0$ as (see~Courcoubetis and
Weber~\cite{Courcoubetis96})
\begin{equation*}
E\left[e^{\theta A_0(t)}\right]=w'e^{\theta r'_{\theta}
t}+we^{\theta r_{\theta} t}~,
\end{equation*}
where $w'=\frac{\lambda
r_{\theta}+\mu(r_{\theta}-P)}{(r_{\theta}-r'_{\theta})(\lambda+\mu)}$,
$w=\frac{-\lambda
r'_{\theta}+\mu(P-r'_{\theta})}{(r_{\theta}-r'_{\theta})(\lambda+\mu)}$,
$r'_{\theta}=\frac{-b-\sqrt{\Delta}}{2\theta}$,
$r_{\theta}=\frac{-b+\sqrt{\Delta}}{2\theta}$,
$b=\lambda+\mu-\theta P$, and $\Delta=b^2+4\mu\theta P$. Since
$w'+w=1$ and $r'_{\theta}\leq r_{\theta}$, it follows that
\begin{equation}
E\left[e^{\theta A_0(t)}\right]\leq e^{\theta r_{\theta}
t}~,\label{eq:mgfb1exp}
\end{equation}
which is the typical approximation of the MGF in the SNC
literature by a single (dominant) exponential. The rate
$r_{\theta}$ corresponds to the \textit{effective bandwidth}; it
is non-decreasing in $\theta$ and satisfies
\begin{equation*}
pP\leq r_{\theta}\leq P~,
\end{equation*}
i.e., $r_{\theta}$ is between the average and the peak rate of a
single MMOO process.

The next result gives a common sample-path bound which will be
used to compute the Standard delay bounds for all three scheduling
algorithms; the result parallels the one from
Theorem~\ref{th:tspb}, which was used for the Martingale bounds.

\begin{theorem}{({\sc{Standard Sample-Path Bound}})}\label{th:spb}
Consider the same hypothesis as in Theorem~\ref{th:tspb}, and in
addition assume that the modulating processes of the sub-flows
start in the steady-state. Then the following sample-path bound
holds for all $0\leq u\leq t$ and $\sigma$
\begin{eqnarray}
&&\hspace{-1cm}\P\left(\sup_{0\leq s<
t-u}\left\{A_1(s,t-u)+A_2(s,t)-C(t-s)\right\}>\sigma\right)\notag\\
&&\hspace{1cm}\leq
\inf_{\left\{\theta:c>r_{\theta}\right\}}Le^{-\theta(C-n_2r_{\theta})u}e^{-\theta\sigma}~,\label{eq:spb}
\end{eqnarray}
where $L=\frac{ce}{c-r_{\theta}}$ and $r_{\theta}$ was defined
prior to Eq.~(\ref{eq:mgfb1exp}).
\end{theorem}

\medskip

The proof proceeds by first discretizing the sample-path event and
then by using standard arguments in SNC and effective bandwidth
theory based on the Union and Chernoff bounds (i.e., the second
and third bounding steps discussed in Section~\ref{sec:pitfall}).
Similar proofs have been given for various sample-path events
(see, e.g.,~\cite{CiBuLi06}).

\proof Fix $0\leq u\leq t$, $\sigma$, and $\theta>0$ such that
$c>r_{\theta}$. Consider the free parameter $\tau_0>0$ for
discretizing the event from Eq.~(\ref{eq:spb}) at the points
$j=\lfloor\frac{t-u-s}{\tau_0}\rfloor$ for all $0\leq s\leq t$.
Using the monotonicity of the arrival processes (e.g.,
$A_1(s,t-u)\leq A_1(t-u-(j+1)\tau_0,t-u)$), we can bound the
probability event from Eq.~(\ref{eq:spb}) by
\begin{eqnarray*}
&&\hspace{-0.5cm}\P\left(\bigcup_{j\geq0}\left\{A_1(t-u-(j+1)\tau_0,t-u)+A_2(t-u-(j+1)\tau_0,t)-C(t-(t-u-j\tau_0))>\sigma\right\}\right)\\
&&\leq\sum_{j\geq1}\P\left(A_1(t-u-j\tau_0,t-u)+A_2(t-u-j\tau_0,t)-C(u+j\tau_0)>-C\tau_0+\sigma\right)\\
&&\leq\sum_{j\geq1}e^{-\theta(C-nr_{\theta})j\tau_0}e^{-\theta(C-n_2r_{\theta})u}e^{\theta
C\tau_0}e^{-\theta\sigma}\\ &&\leq\frac{e^{\theta
C\tau_0}}{\theta(C-nr_{\theta})\tau_0}e^{-\theta(C-n_2r_{\theta})u}e^{-\theta\sigma}~.
\end{eqnarray*}
The derivations relied first on the Union Bound, then on the
Chernoff bound applied to the MGF bound from
Eq.~(\ref{eq:mgfb1exp}), and finally on the inequality
$\sum_{j\geq1}e^{-aj}\leq\frac{1}{a}$ for some $a>0$. Since
$\tau_0$ is a free parameter, the last bound can be optimized with
$\tau_0=\frac{1}{\theta C}$. Finally, taking the minimum over
$\theta$ completes the proof.~\hfill $\Box$

\medskip

Next we list the standard bounds on $A_1(t)$'s virtual delay for
FIFO, SP, EDF, and GPS scheduling. They are obtained alike the
Martingale ones from Eqs.~(\ref{eq:fifodb}), (\ref{eq:spdb}), and
(\ref{eq:gpsdb}), except that the sample-path bound from
Theorem~\ref{th:tspb} is replaced by the one from
Theorem~\ref{th:spb}:
\begin{eqnarray}
\hspace{-0.55cm}\textrm{\textbf{FIFO}}&:&\ldots\leq
\inf_{\left\{\theta:c>r_{\theta}\right\}}Le^{-\theta
Cd}\label{eq:fifodbStandard}\\
\hspace{-0.55cm}\textrm{\textbf{SP}}&:&\ldots\leq
\inf_{\left\{\theta:c>r_{\theta}\right\}}Le^{-\theta(C-n_2r_{\theta})d}\label{eq:spdbStandard}\\
\hspace{-0.55cm}\textrm{\textbf{EDF$^1$}}&:&\ldots\leq
\inf_{\left\{\theta:c>r_{\theta}\right\}}Le^{\theta
n_2r_{\theta}\min\left\{d_1^*-d_2^*,d\right\}}e^{-\theta
Cd}\label{eq:edf1dbStandard}\\
\hspace{-0.55cm}\textrm{\textbf{EDF$^2$}}&:&\ldots\leq
\inf_{\left\{\theta:c>r_{\theta}\right\}}Le^{\theta\left(C-
n_1r_{\theta}\right)\left(d_1^*-d_2^*\right)}e^{-\theta
Cd}+\inf_{\left\{\theta:c'>r_{\theta}\right\}}L'e^{-\theta Cd}
\label{eq:edf2dbStandard}\\
\hspace{-0.55cm}\textrm{\textbf{GPS}}&:&\ldots\leq
\inf_{\left\{\theta:\phi_1C>n_1r_{\theta}\right\}}Le^{-\theta
\phi_1Cd}\label{eq:gpsdbStandard}
\end{eqnarray}
where $L$ is given in Theorem~\ref{th:spb} for FIFO and SP, and
$L=\frac{\phi_1 C}{\phi_1C-n_1r_{\theta}}$ for GPS. We mention
that \textrm{\textbf{EDF$^1$}} corresponds to the case
$d_1^*>d_2^*$ (see the Martingale bound from
Eq.~(\ref{eq:edfdb1})), and \textrm{\textbf{EDF$^2$}} to the
complementary case (see Eq.~(\ref{eq:edfdb2})). Moreover, for
\textrm{\textbf{EDF$^2$}}, $c'$ is the rescaled value of $c$ (as
in the Martingale bound from Eq.~(\ref{eq:edfdb2})), whereas $L'$
is defined like $L$ but with $c$ replaced by $c'$. We also point
out that the GPS bound is roughly the same as the one computed by
Zhang~\et~\cite{Zhang94}, for more general arrival classes and
without using the service process concept (which was still to be
properly formalized by Cruz~\cite{Cruz95} a year after).

\subsection{Asymptotic Decay Rates Comparison}
Here we compare the asymptotic decay rates (denoted by $\eta$) of
the Martingale and Standard bounds on $\P(W_1(t)>d)$, which can be
isolated by taking the limit
\begin{equation*}
\eta:=-\lim_{d\rightarrow\infty}\frac{\log{\P\left(W_1(t)>d\right)}}{d}~.
\end{equation*}

\begin{table}[h]\renewcommand{\arraystretch}{2}\addtolength{\tabcolsep}{-1pt}
\begin{center}
\begin{tabular}{c| c | c }
Delay Bounds~/~Scheduling&Martingale & Standard \\
\hline FIFO, EDF & $\gamma C$ & $\theta^*C$\\
\hline SP & $\gamma C_1$ & $\theta^*\left(C-n_2r_{\theta^*}\right)$\\
\hline GPS & $\gamma\phi_1C$ & $\theta^*\phi_1C$\\
\hline
\end{tabular}
\end{center}
\caption{Asymptotic decay rates for $\P\left(W_1(t)>d\right)$; for
FIFO, EDF, and GPS: $\gamma=\theta^*$} \label{tb:adr}
\end{table}

Table~\ref{tb:adr} lists the asymptotic decay rates for FIFO, SP,
EDF, and GPS scheduling. For the Standard bounds, the $\theta^*$'s
are the solutions to the optimization problems from
Eqs.~(\ref{eq:fifodbStandard})-(\ref{eq:gpsdbStandard}),
respectively. Next we discuss the bounds by grouping them into two
groups.

\subsubsection{FIFO, EDF, and GPS}
We only focus on FIFO; the other two are analogous.

For the Standard FIFO bound, $\theta^*$ is the unique solution of
the equation
\begin{equation}
r_{\theta}=c~,\label{eq:efb}
\end{equation}
where $r_{\theta}$ is the effective bandwidth from
Eq.~(\ref{eq:mgfb1exp}). This is a standard result in the
effective bandwidth literature (see, e.g., Glynn and
Whitt~\cite{GlynnWhitt1994} or Chang~\cite{Book-Chang}, p. 291).

Next we show that $\theta^*=\gamma$, i.e., the asymptotic decay
rates are equal. To this end, we first recall
from~\cite{Courcoubetis96} that $r_{\theta}$ is derived by solving
the eigenvalue problem
\begin{equation}
{\mathbf{Q}}_{\theta}{\mathbf{x}}=\zeta_{\theta}{\mathbf{I}}{\mathbf{x}}\label{eq:eigr}
\end{equation}
where ${\mathbf{x}}$ stands for the eigenvector and
\begin{equation*}
{\mathbf{Q}}_{\theta}:=\left(\begin{array}{cc}-\mu&\mu\\\lambda&-\lambda+P\theta\end{array}\right),~{\mathbf{I}}:=\left(\begin{array}{cc}1&0\\0&1\end{array}\right)~,
\end{equation*}
for some $\theta>0$. If $\zeta_{\theta}$ denotes the spectral
radius of ${\mathbf{Q}}_{\theta}$ (corresponding to $\omega_2$ in
\cite{Courcoubetis96}, Section 3), then the effective bandwidth is
defined as $r_{\theta}=\frac{\zeta_{\theta}}{\theta}$ (see
Eq.~(\ref{eq:mgfb1exp})).

In turn, $\gamma$ from the Martingale bound satisfies the
generalized eigenvalue problem (see~\cite{Palmowski96})
\begin{equation}
{\mathbf{Q}}_0{\mathbf{y}}=\gamma{\mathbf{D}}{\mathbf{y}}~,\label{eq:eiggamma}
\end{equation}
where ${\mathbf{y}}$ stands for the eigenvector and
\begin{equation*}
{\mathbf{D}}:=\left(\begin{array}{cc}c&0\\0&c-p\end{array}\right)~.
\end{equation*}
Using the positivity of $\gamma$ (see Theorem~\ref{th:tspb}) and
applying the Separation Lemma from Sonneveld~\cite{Sonneveld04},
it follows that $\gamma$ is the spectral radius. Furthermore, by
relating the eigenvalue problems from Eqs.~(\ref{eq:eigr}) and
(\ref{eq:eiggamma}) through the equality
\begin{equation*}
{\mathbf{Q}}_{\gamma}-c\gamma{\mathbf{I}}={\mathbf{Q}}_0-\gamma{\mathbf{D}}~,
\end{equation*}
it follows that $c\gamma=\zeta_{\gamma}$. Finally, since
$\theta^*$ is the unique solution of Eq.~(\ref{eq:efb}), whereas
$r_{\theta^*}=\frac{\zeta_{\theta^*}}{\theta^*}$ by definition, it
follows that $\theta^*=\gamma$. This result is particularly
important as it illustrates an explicit solution for the effective
bandwidth equation from Eq.~(\ref{eq:efb}).

\subsubsection{SP}
Unlike in the FIFO case, there is no immediate explicit solution
for the optimal value $\theta^*$. Because $c=r_{\gamma}$, as shown
above, we can observe that
\begin{equation*}
\gamma C_1=\gamma\left(C-n_2r_{\gamma}\right)~,
\end{equation*}
which indicates a similarity in the two asymptotic decay rates.
Unfortunately, due the form of the optimization problem from
Eq.~(\ref{eq:spdbStandard}), we will solely resort to numerical
comparisons between the Martingale and Standard bounds.

In conclusion, in large buffer (or, here, delay) asymptotic
regimes, FIFO, EDF, and GPS exhibit the same tail behavior for
both the Standard and Martingale bounds; according to numerical
results, the same holds for SP (see Section~\ref{sec:simus}). Next
we consider many-sources asymptotics, which are also commonly
discussed in the literature (e.g., Mazumdar~\cite{Mazumdar09}).

\subsection{Many-Sources Asymptotics Comparison}
We consider the following scaling scheme: the total number of
flows $n$ is scaled up, whereas the rest of the parameters, i.e.,
the utilization factor $\rho$, the per-flow rate $r_0=pP$, and the
per-flow capacity $c$ remain unchanged.

Let us first observe that the factors $K$ (defined in
Theorem~\ref{th:tspb}) and $L$ (defined in Theorem~\ref{th:spb})
from the two sets of bounds satisfy
\begin{equation*}
K<1~\textrm{and}~L>1~.
\end{equation*}
The second property is immediate. In turn, for the factor $K$,
note that $\frac{\rho-p}{1-p}<1$ and the functions
$f(x):=\frac{\rho-x}{1-x}$ and $g(x):=\frac{x-\rho}{\rho}$ are
non-increasing on $x\in\left(0,\rho\right)$. Thus, by the
composition of the power function (defined on the interval
$(0,1)$) and the exponential function, the function
$h(x):=\left(\frac{\rho-x}{1-x}\right)^{\frac{x-\rho}{\rho}}$ is
non-increasing. Moreover, $\lim_{x\downarrow0}h(x)=\rho^{-1}$, and
thus $K<1$.

\begin{table}[h]\renewcommand{\arraystretch}{2}\addtolength{\tabcolsep}{-1pt}
\begin{center}
\begin{tabular}{c| c | c }
Delay Bounds~/~Scheduling&Martingale & Standard \\
\hline FIFO, SP, EDF, GPS & ${\mathcal{O}}\left(e^{-\alpha n}e^{-\eta d n}\right)$ & ${\mathcal{O}}\left(e^{-\eta d n}\right)$\\
\hline
\end{tabular}
\end{center}
\caption{Scaling laws for the Martingale and Standard bounds on
$\P\left(W_1(t)>d\right)$ in the total number of flows
$n=n_1+n_2$; $\alpha$ and $\eta$ are invariant to $n$}
\label{tb:slaws}
\end{table}

Table~\ref{tb:slaws} illustrates the scaling laws of the
Martingale and Standard delay bounds for the four scheduling
algorithms. The factors $\alpha>0$ and $\eta>0$ are invariant to
$n$ and can be fitted for each individual case; e.g., in the case
of FIFO, $\alpha=-\log{K}$ and $\eta=\gamma c$. We remark that all
pairs of bounds have the same \textit{asymptotic decay rate}
$\eta$. The critical observation is that, unlike the Standards
bounds, the Martingale bounds have an additional factor
$e^{-\alpha n}$ decaying exponentially with $n$. This scaling
behavior was indicated by
Choudhury~\et~\cite{choudhury96squeezing} by numerical
evaluations. We point out that~\cite{choudhury96squeezing} further
indicated an additional factor $\beta>0$, invariant to $n$, which
is however not captured by the Martingale bounds.

\subsection{Numerical Comparisons}\label{sec:simus}
In this section we compare the Martingale and Standard bounds
against simulations and also in asymptotic regimes. The parameters
of a single MMOO source are $\lambda=0.5$, $\mu=0.1$, and $P=1$
(the average `Off' period is five fold the average `On' period).

\subsubsection{Bounds vs. Simulations}
We consider two utilization levels ($\rho=0.75$ and $\rho=0.9$),
and two degrees of multiplexing ($n_1=n_2=5$ and $n_1=n_2=10$).
The packet sizes in a packet-level simulator are set to $1$;
fractional packet sizes are additionally set when the dwell times
in the states of the Markov process from Figure~\ref{fig:nonoff}
are not integers. The simulator measures the delays of the through
flow's first $10^7$ packets, and it discards the first $10^6$. For
numerical confidence, $100$ independent simulations are being run
and the results are presented as box-plots.

For the soundness of the comparisons against simulations, it is
important to remark that the delay analysis so far concerned the
\textit{virtual delay process} $W_1(t)$, which corresponds to the
delay of a through flow's infinitesimal unit, should it depart, or
equivalently arrive, at time $t$; more concretely, we note that
the bounds computed with SNC on virtual delays are identical,
should they concern a virtual arrival or departure unit. In the
packet level simulator, however, it is the \textit{packet delay
process} which is being measured, and which is denoted here by
$\widetilde{W}_1(n)$ (the index `$n$' corresponds to the packet
number for the through flow). Therefore, one has to properly
perform a suitable change of probability measures in order to
provide meaningful numerical comparisons.

We follow a Palm calculus argument and relate the measure of the
virtual delay process to that of the packet delay process (see
Shakkottai and Srikant~\cite{Shakkottai00}). For convenience, we
work in reversed time and focus on time $0$ where steady-state is
assumed to be reached. Denoting $W_1:=W_1(0)$, we can write by
conditioning
\begin{eqnarray}
\P\left(W_1>d\right)&=&\P\left(W_1>d\mid
a_1(0)>0\right)\P\left(a_1(0)>0\right)+\P\left(W_1>d\mid a_1(0)=0\right)\P\left(a_1(0)=0\right)\notag\\
&\geq&\P\left(W_1>d\mid
a_1(0)>0\right)\P\left(a_1(0)>0\right)\notag\\
&=&\P\left(\widetilde{W}_1>d\right)\P\left(a_1(0)>0\right)~,\label{eq:palm}
\end{eqnarray}
where $a_1(0)$ denotes the instantaneous arrivals of the through
flow at time $0$, and $\widetilde{W}_1$ denotes the steady-state
packet delay process of the through flow. Note that for the
inequality we eliminated the second term in the sum above.
\begin{figure}[h]
\vspace{-0cm}
\shortstack{\hspace{-0.165cm}
\includegraphics[width=0.522\linewidth,keepaspectratio]{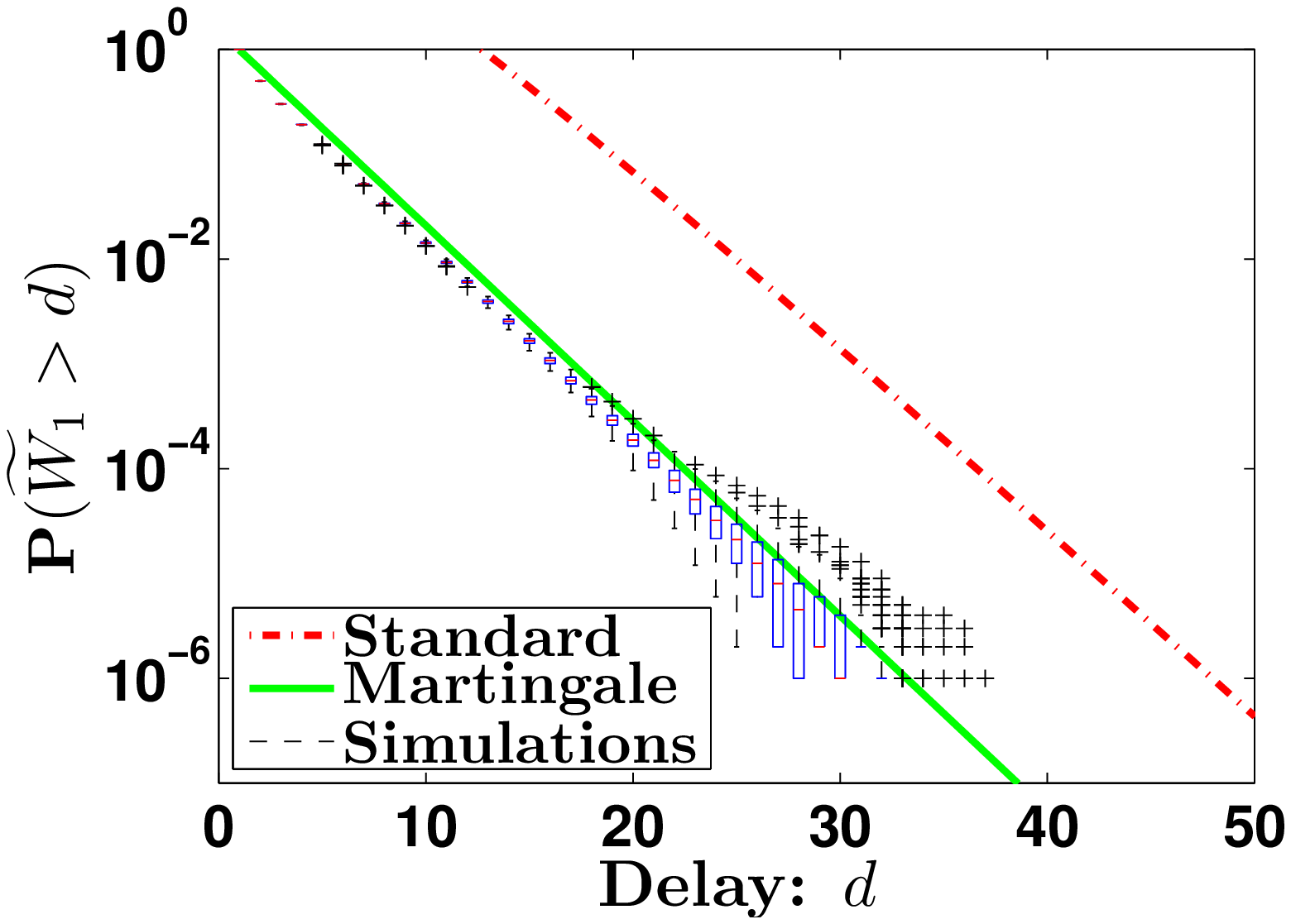}
\\
{\footnotesize (a) $\rho=0.75$ ($n_1=n_2=5$)} }
\shortstack{\hspace{-0.5cm}
\includegraphics[width=0.522\linewidth,keepaspectratio]{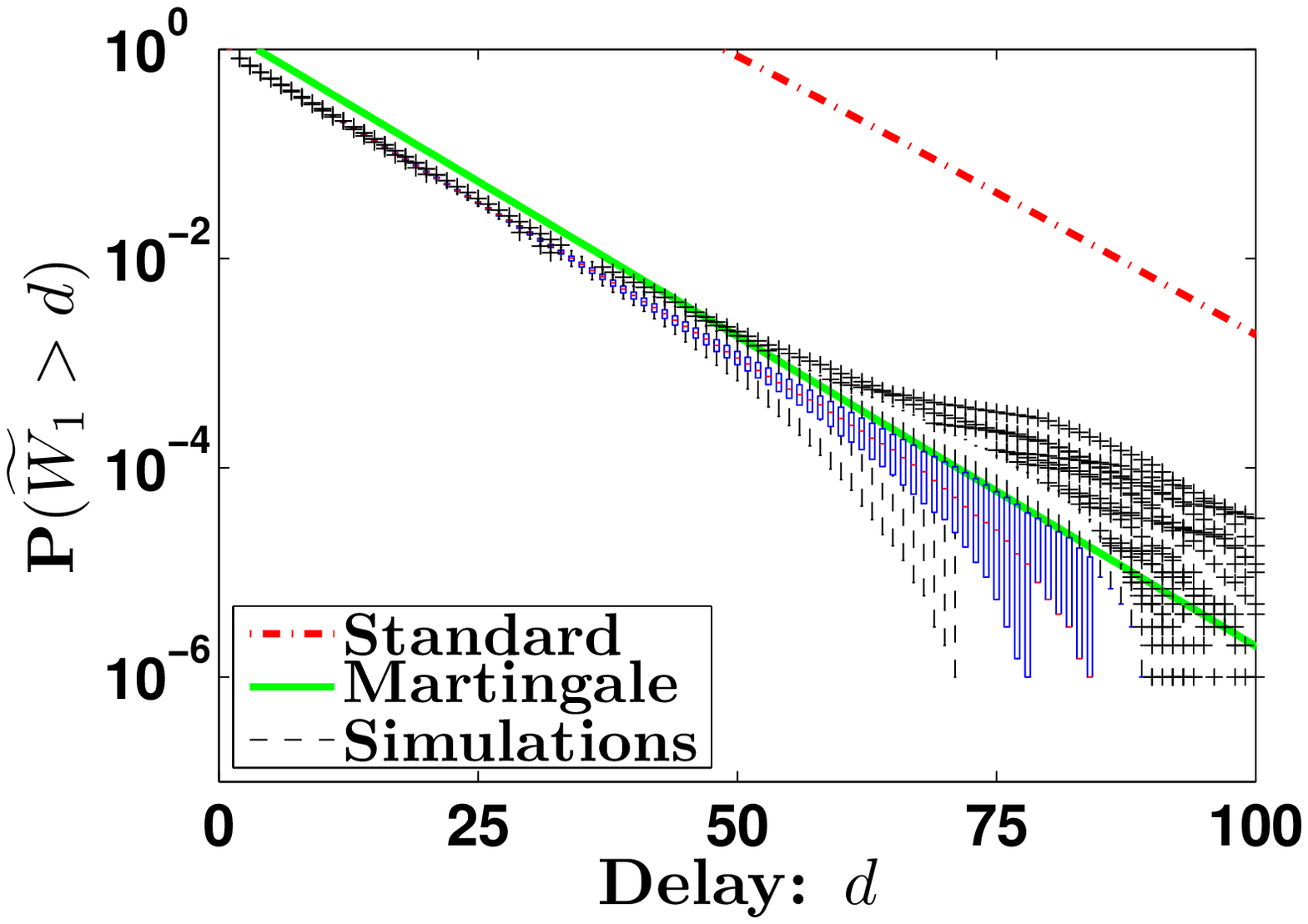}
\\
{\footnotesize (b) $\rho=0.90$} ($n_1=n_2=5$)}
\shortstack{\hspace{-0.165cm}
\includegraphics[width=0.522\linewidth,keepaspectratio]{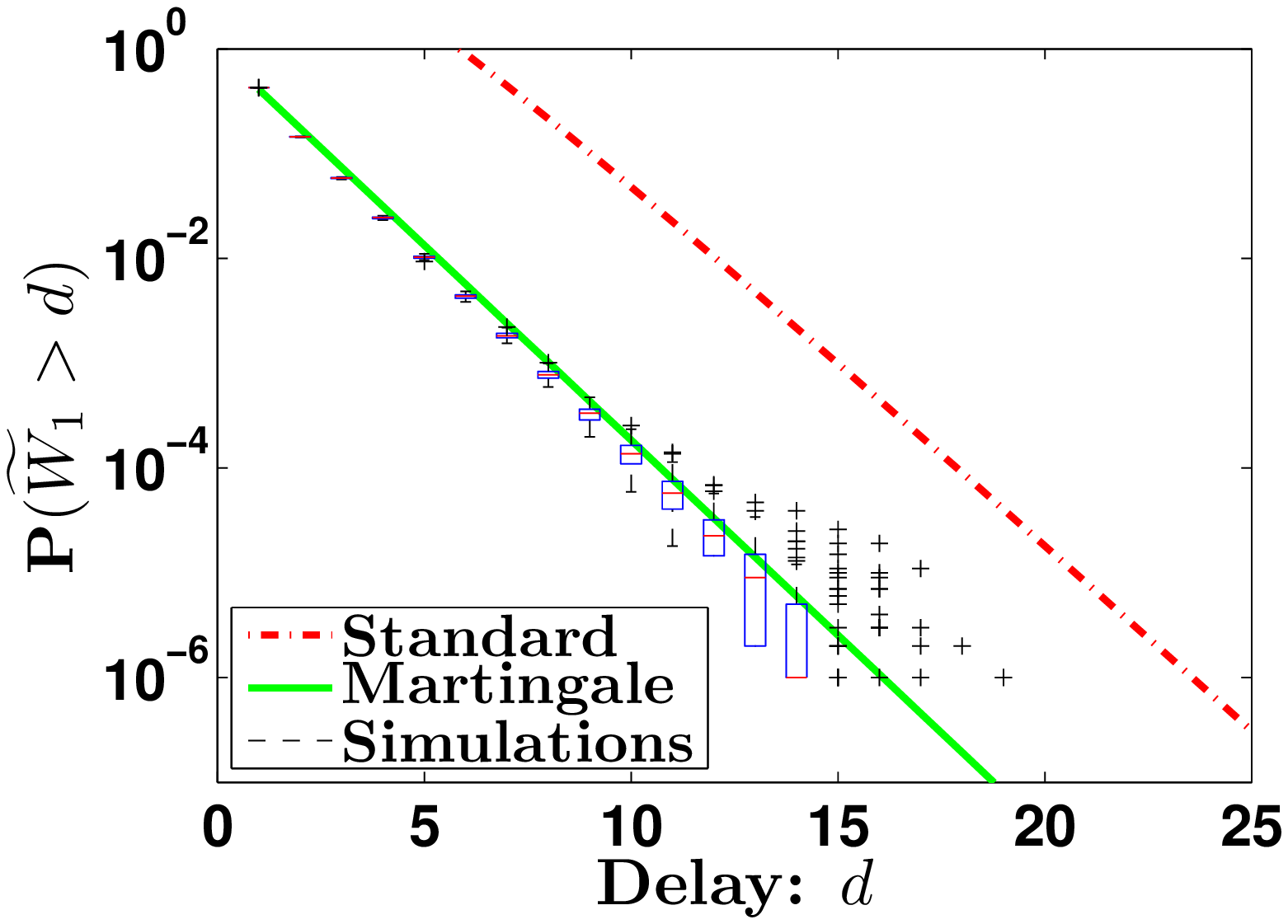}
\\
{\footnotesize (c) $\rho=0.75$ ($n_1=n_2=10$)} }
\shortstack{\hspace{-0.5cm}
\includegraphics[width=0.522\linewidth,keepaspectratio]{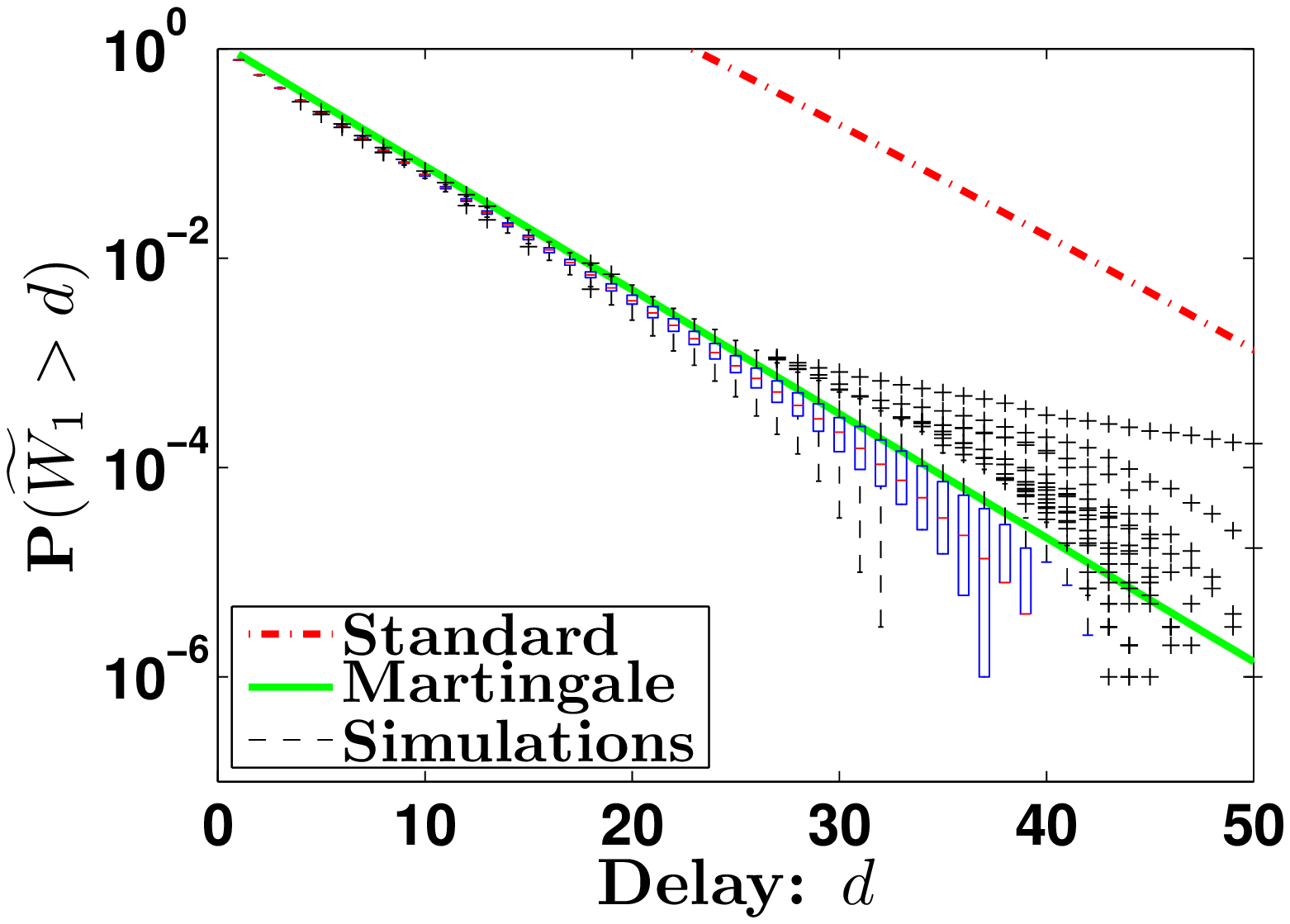}
\\
{\footnotesize (d) $\rho=0.90$} ($n_1=n_2=10$)}
 \caption{FIFO delay
bounds} \label{fig:fifo}
\end{figure}

Therefore,
\begin{equation}
\P\left(\widetilde{W}_1>d\right)\leq\frac{1}{1-(1-p)^n}\P\left(W_1>d\right)~.\label{eq:corbounds}
\end{equation}
(Recall that $p$ is the steady-state `On' probability of the MMOO
process from Figure~\ref{fig:onoff}.)

Below we compare the distribution of $\widetilde{W}_1$ against the
one of the measured (simulated) delay process. All the Martingale
and Standard bounds which we compute for FIFO, SP, EDF, and GPS
scheduling are scaled up by the additional prefactor from
Eq.~(\ref{eq:corbounds}) needed for the change of measure. We note
that this scaling is conservative because of the inequality from
Eq.~(\ref{eq:palm}).

Figure~\ref{fig:fifo} illustrates the comparisons for FIFO
scheduling (recall Eqs.~(\ref{eq:fifodb}) and
(\ref{eq:fifodbStandard}) for the Martingale and Standard delay
bounds, which are scaled as in Eq.~(\ref{eq:corbounds})); the
$y$-axis uses a log scale. The irregular tail behavior (including
the presence of many outliers\footnote{Outliers are depicted in
the box-plots with the `+' symbol; on each box, the central mark
is the median, and the edges of the box are the 25th and 75th
percentiles.}\footnote{The long stretch of the box-plots and the
presence of many outliers is caused by the choice of $10^7$
arrivals, in order to illustrate the need for very long simulation
runs (e.g., $10^8$ arrivals, in which case the box-plots would
significantly shrink and most outliers would disappear).}at
$\rho=90\%$) of the box-plots is due to the restriction of the
simulation runs to $10^7$ packets. All four scenarios,
corresponding to various utilizations and multiplexing, clearly
indicate that the Standard bounds are very loose, as they
overestimate the simulation results by a factor of roughly
$10^{2}$ at $75\%$ utilization (see (a) and (c)), and even
$10^{3}$ at $90\%$ utilization (see (b) and (d)). In turn, the
Martingale bounds are reasonably accurate. We suspect that the
slight loss of accuracy for $n_1=n_2=5$ (in (a) and (b)) stems
from the conservative change of measure from Eq.~(\ref{eq:palm});
indeed, at $n_1=n_2=10$ (in (c) and (d)) one can notice an
increase in accuracy due to the lesser role played by the change
of measure prefactor from Eq.~(\ref{eq:corbounds}).

\begin{figure}[h]
\vspace{-0cm}
\shortstack{\hspace{-0.165cm}
\includegraphics[width=0.5205\linewidth,keepaspectratio]{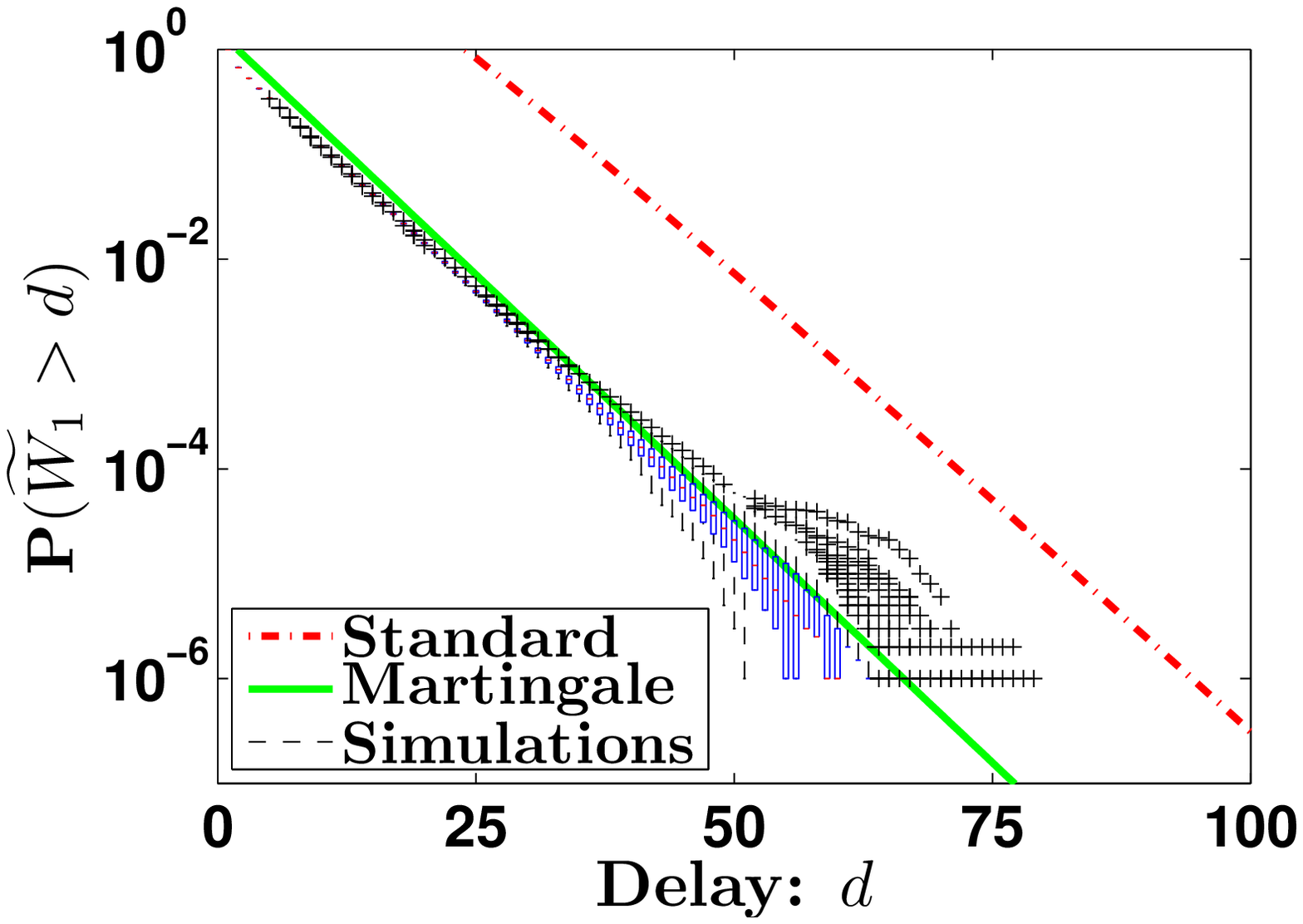}
\\
{\footnotesize (a) $\rho=0.75$ ($n_1=n_2=5$)} }
\shortstack{\hspace{-0.495cm}
\includegraphics[width=0.5205\linewidth,keepaspectratio]{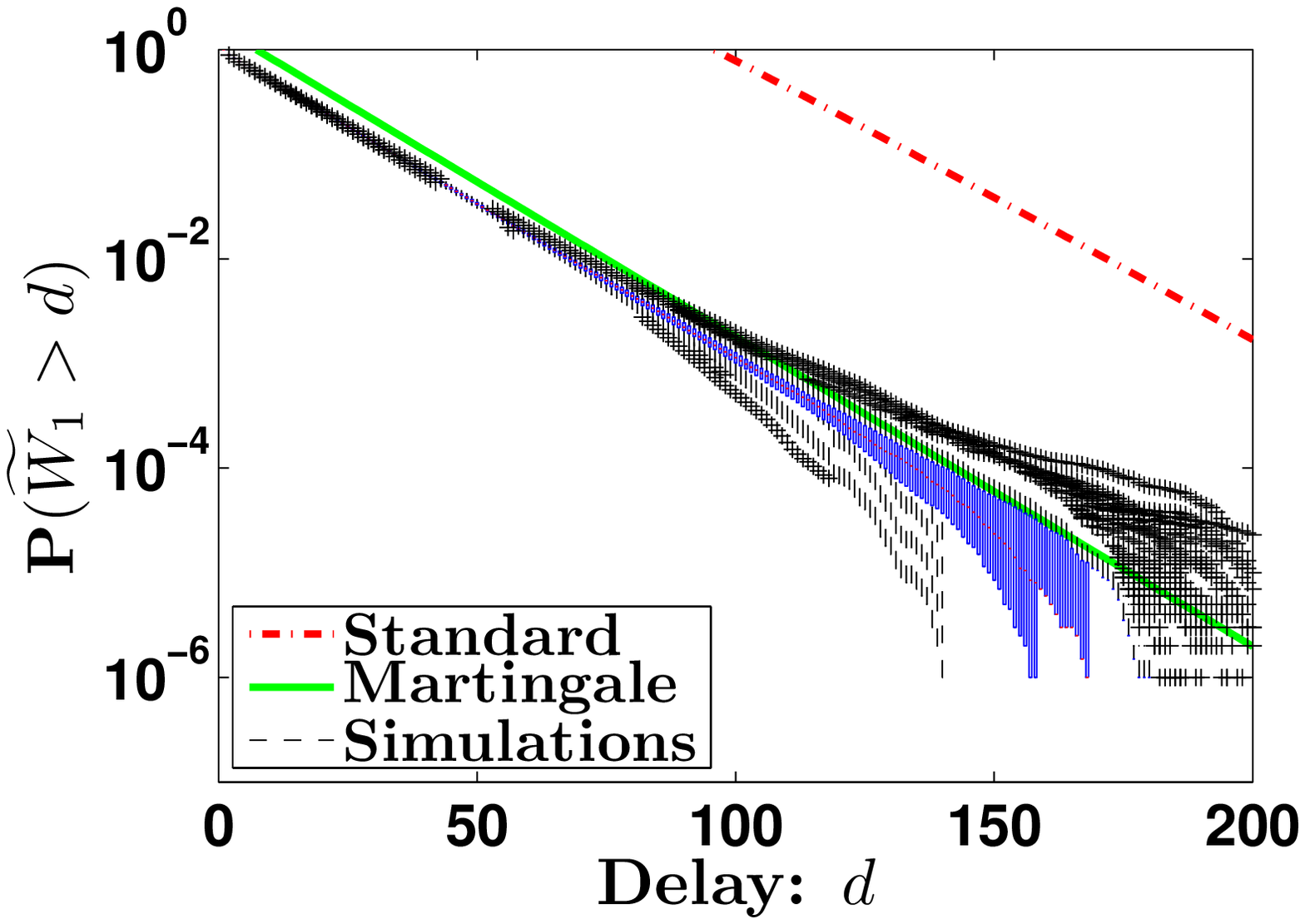}
\\
{\footnotesize (b) $\rho=0.90$} ($n_1=n_2=5$)}
\shortstack{\hspace{-0.165cm}
\includegraphics[width=0.522\linewidth,keepaspectratio]{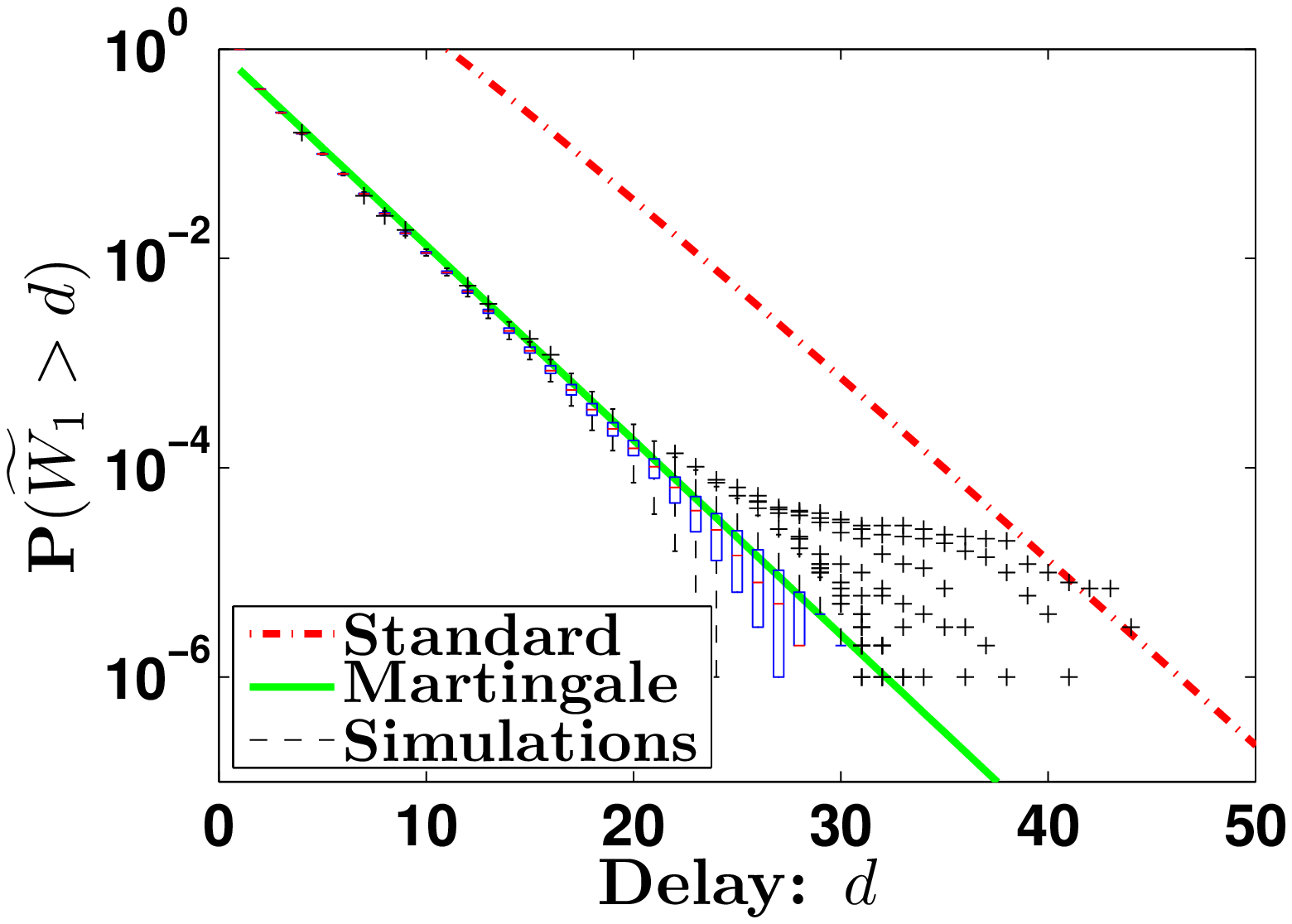}
\\
{\footnotesize (c) $\rho=0.75$ ($n_1=n_2=10$)} }
\shortstack{\hspace{-0.5cm}
\includegraphics[width=0.522\linewidth,keepaspectratio]{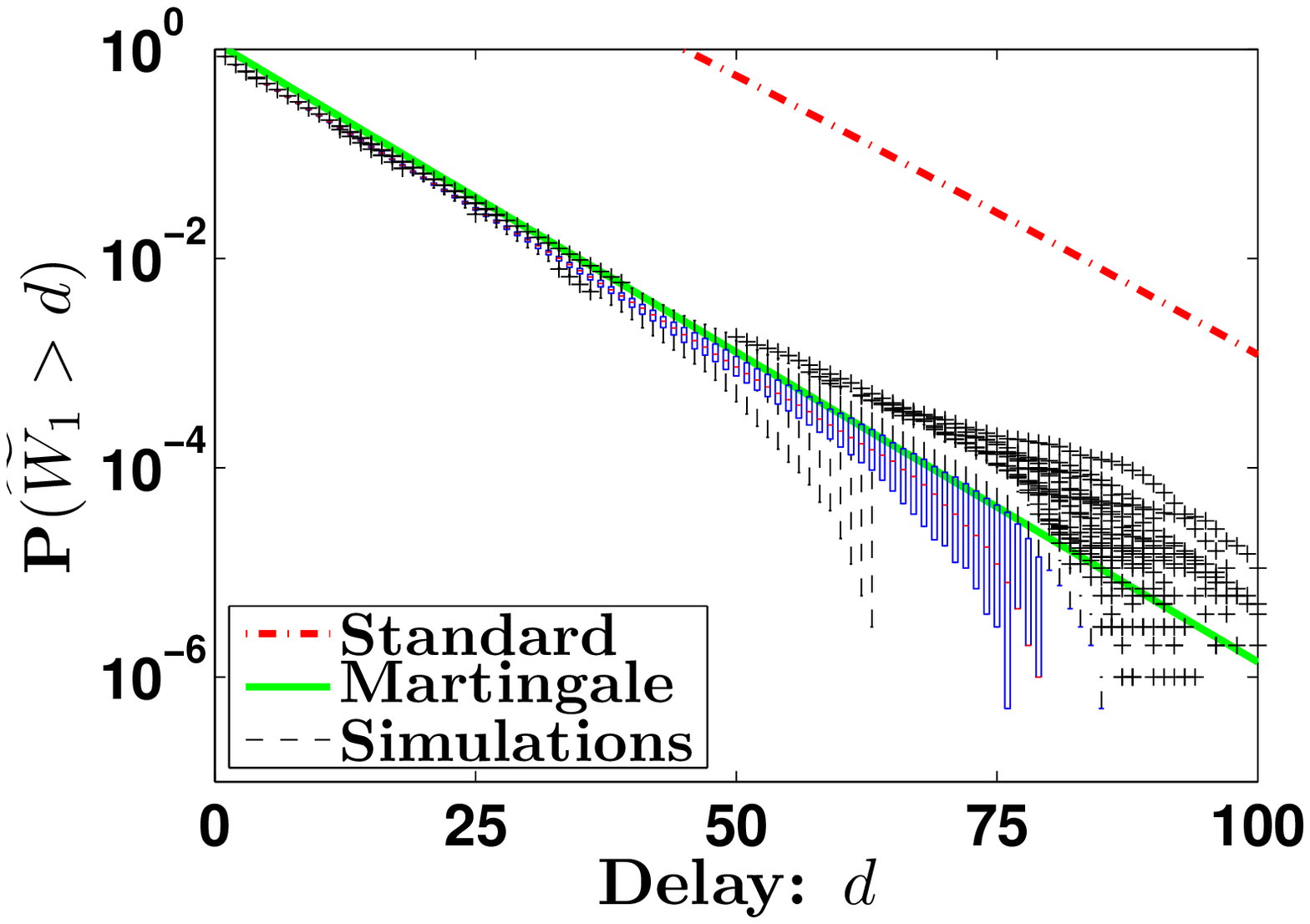}
\\
{\footnotesize (d) $\rho=0.90$} ($n_1=n_2=10$)}
 \caption{SP delay
bounds} \label{fig:sp}
\end{figure}

The same observations hold for SP scheduling, as indicated by
Figure~\ref{fig:sp}; recall the Martingale and Standard delay
bounds from Eq.~(\ref{eq:spdb}) and (\ref{eq:spdbStandard}),
respectively, which are again scaled as in
Eq.~(\ref{eq:corbounds}). Moreover, note that the SP delays
increase roughly by a factor of $2$ relative to the FIFO delays,
due to the setting $n_1=n_2$; the same factor predominates at
higher multiplexing regimes as well. While this is an indication
that scheduling can matter, we refer to Section~\ref{sec:useful}
for a specific asymptotic scenario (i.e., with many flows) in
which scheduling does \textit{not} matter.

\begin{figure}[h]
\vspace{-0cm}
\shortstack{\hspace{-0.165cm}
\includegraphics[width=0.522\linewidth,keepaspectratio]{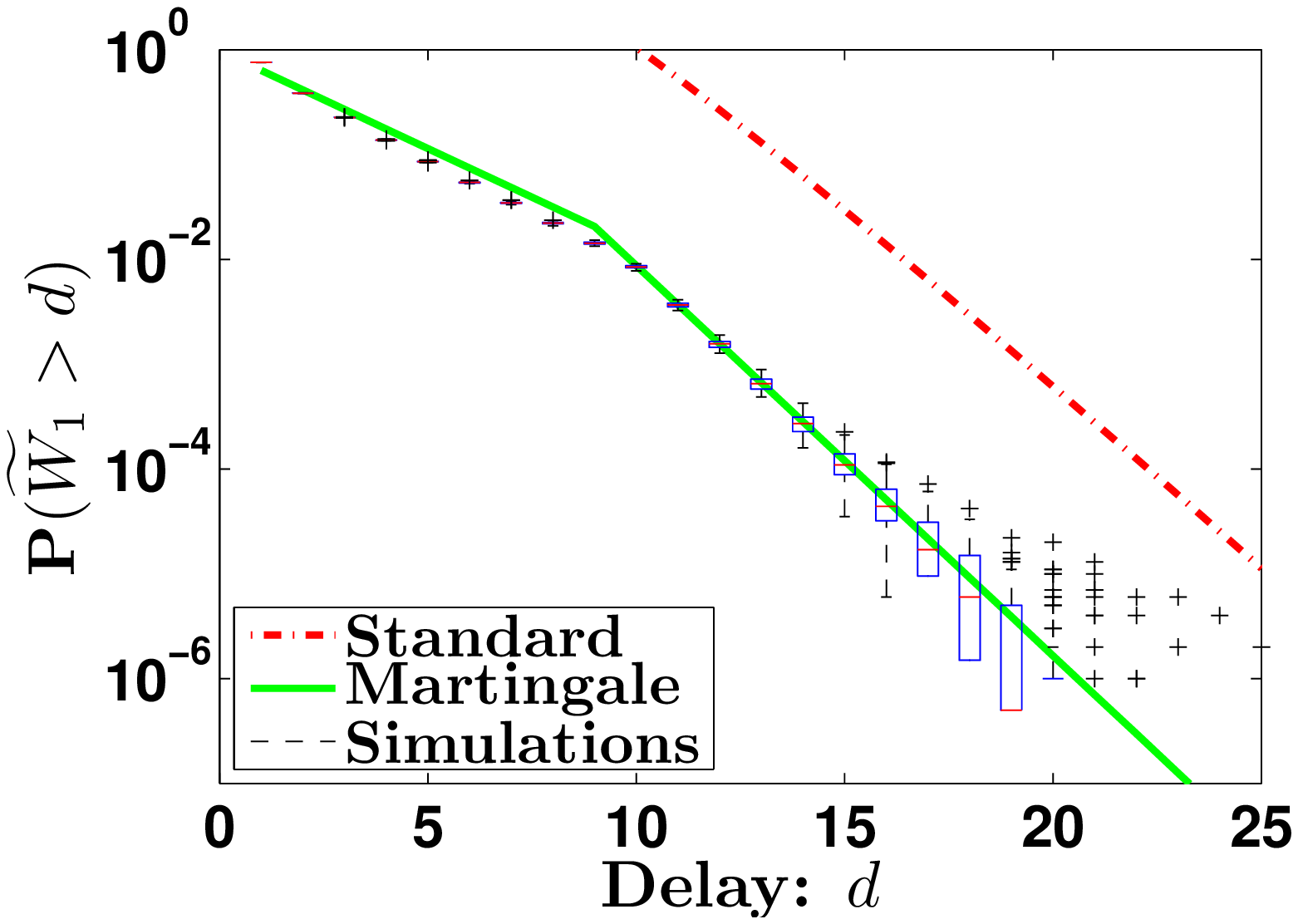}
\\
{\footnotesize (a) $d_1^*=10$, $d_2^*=1$} }
\shortstack{\hspace{-0.495cm}
\includegraphics[width=0.522\linewidth,keepaspectratio]{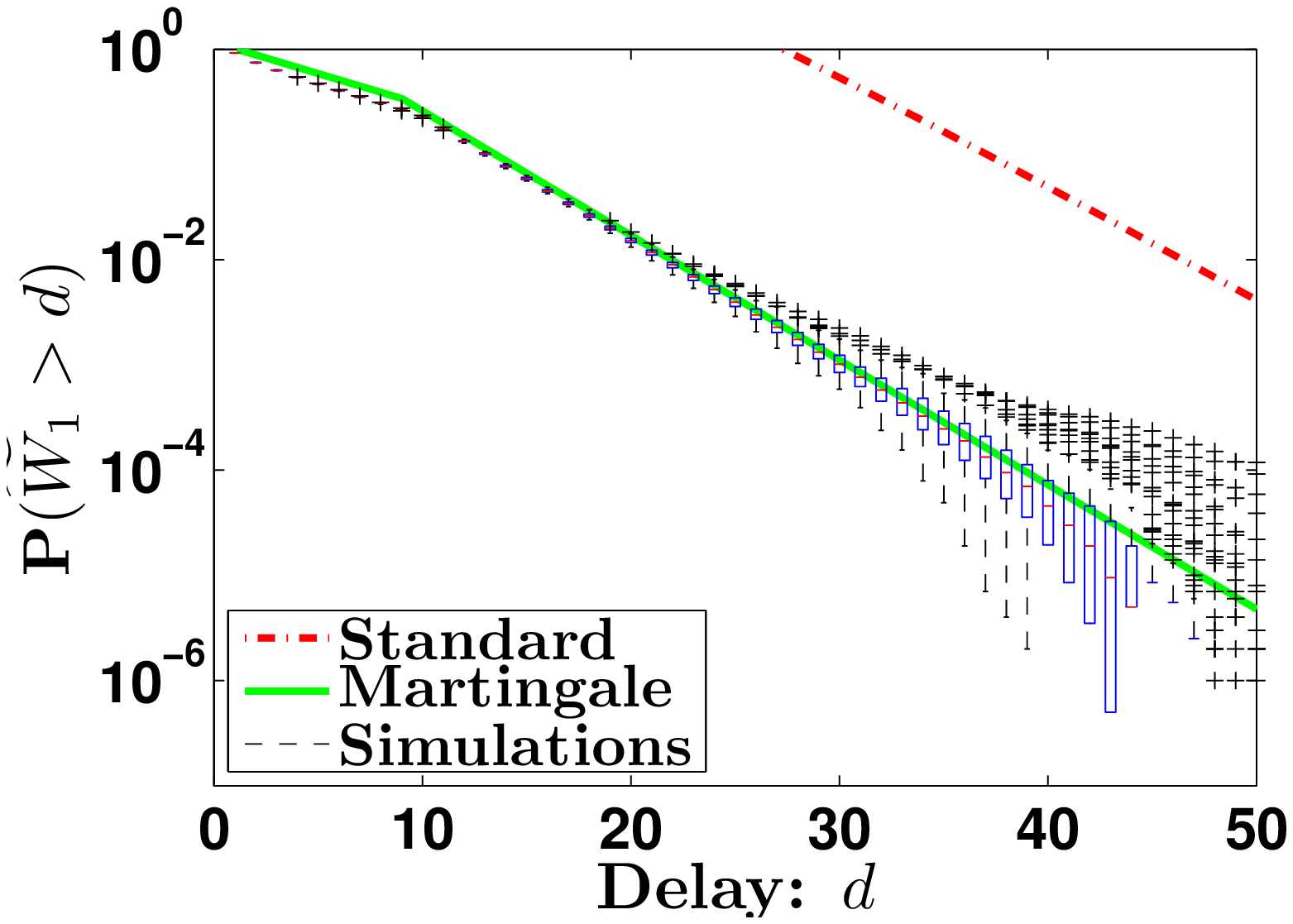}
\\
{\footnotesize (b) $d_1^*=10$, $d_2^*=1$}}
\shortstack{\hspace{-0.165cm}
\includegraphics[width=0.522\linewidth,keepaspectratio]{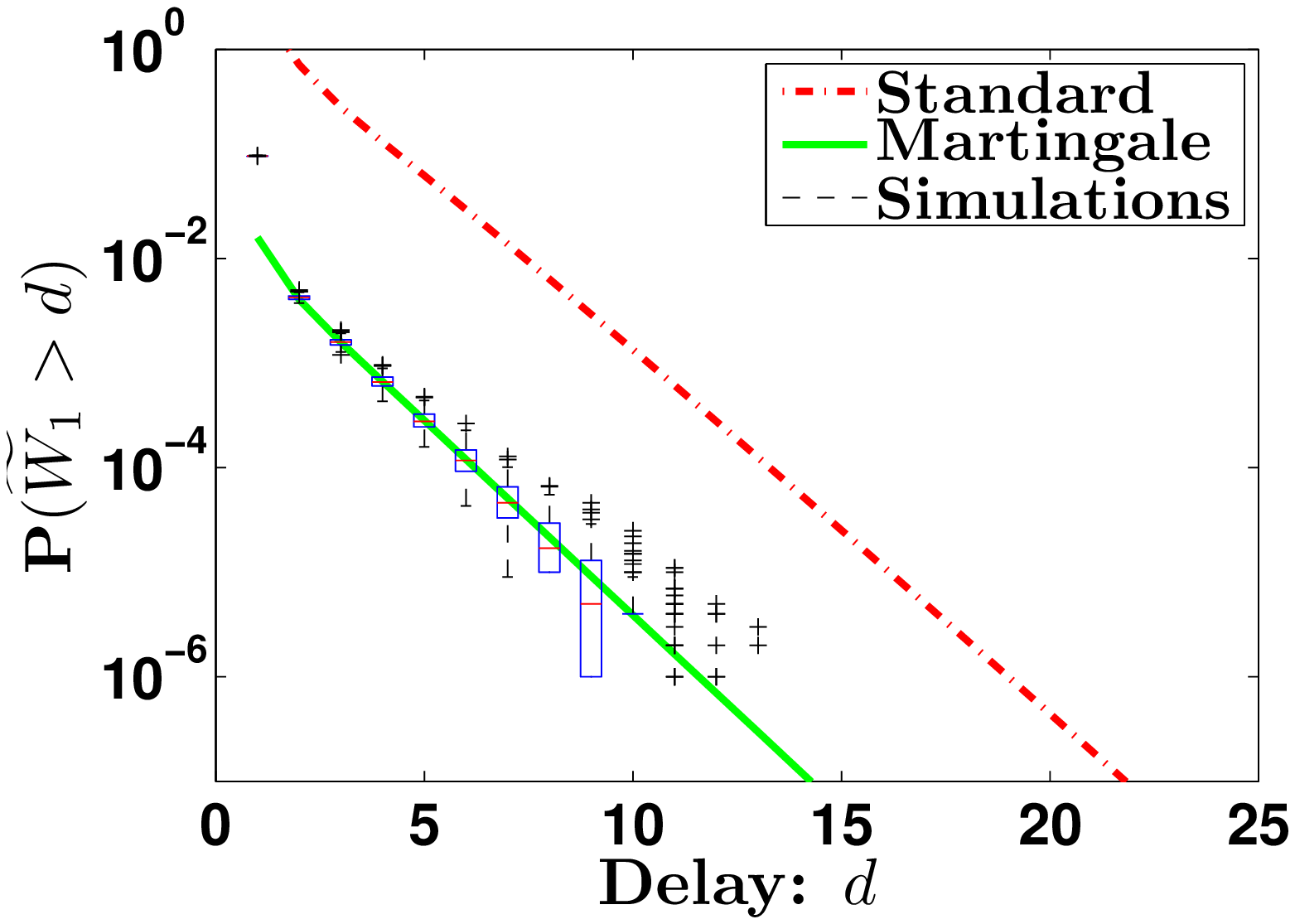}
\\
{\footnotesize (c) $d_1^*=1$, $d_2^*=10$} }
\shortstack{\hspace{-0.495cm}
\includegraphics[width=0.522\linewidth,keepaspectratio]{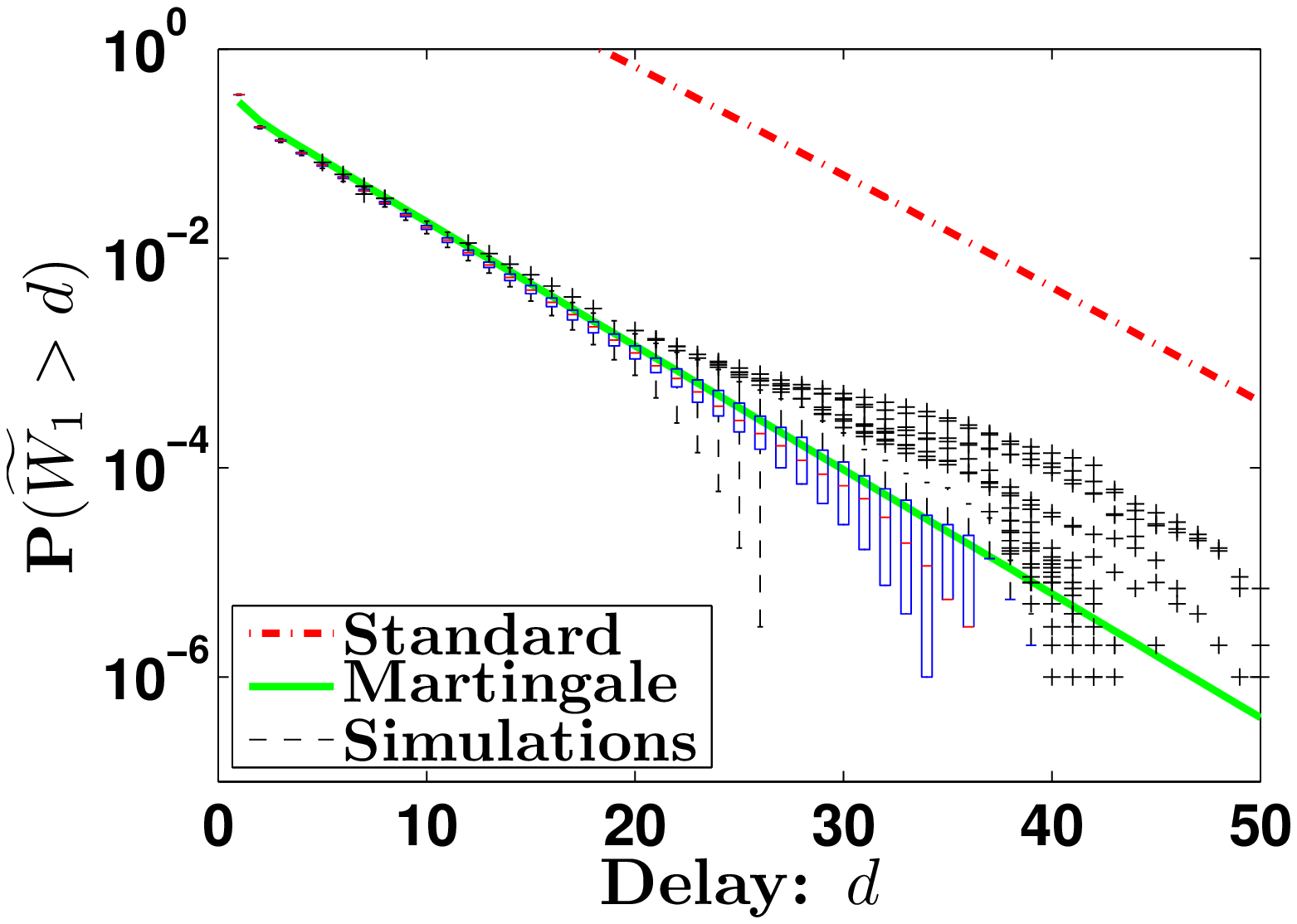}
\\
{\footnotesize (d) $d_1^*=1$, $d_2^*=10$}}
 \caption{EDF delay
bounds ($n_1=n_2=10$, $\rho=75\%$ in (a) and (c), and $\rho=90\%$
in (b) and (d))} \label{fig:edf}
\end{figure}

The tightness of the Martingale bounds, in contrast to the
looseness of Standard bounds, further holds in the case of EDF
scheduling, for both cases (i.e., $d_1^*>d_2^*$ and
$d_1^*<d_2^*$), as illustrated in Figure~\ref{fig:edf}; recall the
Martingale and Standard delay bounds from
Eqs.~(\ref{eq:edfdb1})-(\ref{eq:edfdb2}) and
Eqs.~(\ref{eq:edf1dbStandard})-(\ref{eq:edf2dbStandard}),
respectively. Note that the bending of the curves, e.g., in (a),
is due to the choice of $d_1^*$ and $d_2^*$: the bounds behave
like the SP bounds for $d\leq d_1^*-d_2^*$, and asymptotically
like the FIFO bounds thereafter.

The previous numerical illustrations conceivably indicate the
tightness of the `proprietary' SNC bounding step from
Eq.~(\ref{eq:genb}), in the case of FIFO, SP, and EDF scheduling.
More concretely, the state-of-the-art service processes for these
scheduling algorithms (see Eqs. (\ref{eq:fifosc}),
(\ref{eq:sprocsp}), and (\ref{eq:edfsc})) appear to be tight, as
long as they are used in conjunction with the `right' tools from
probability theory. The same illustrations also confirm the
pitfall identified in Section~\ref{sec:pitfall}, concerning the
potential looseness of the Union Bound in the case of bursty
processes.

\begin{figure}[h]
\vspace{-0cm}
\shortstack{\hspace{-0.165cm}
\includegraphics[width=0.522\linewidth,keepaspectratio]{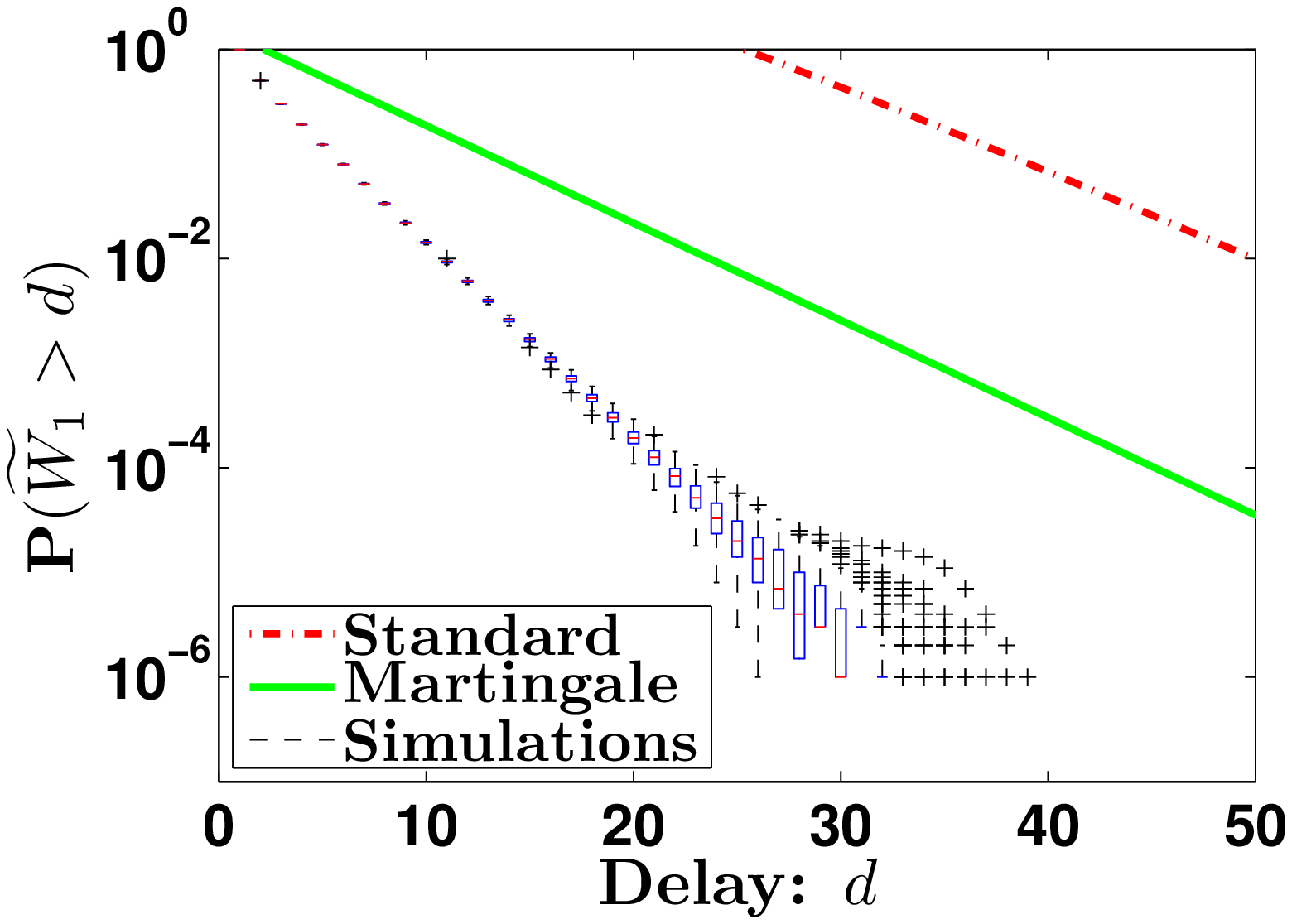}
\\
{\footnotesize (a) $\rho=0.75$ ($n_1=n_2=5$)} }
\shortstack{\hspace{-0.495cm}
\includegraphics[width=0.522\linewidth,keepaspectratio]{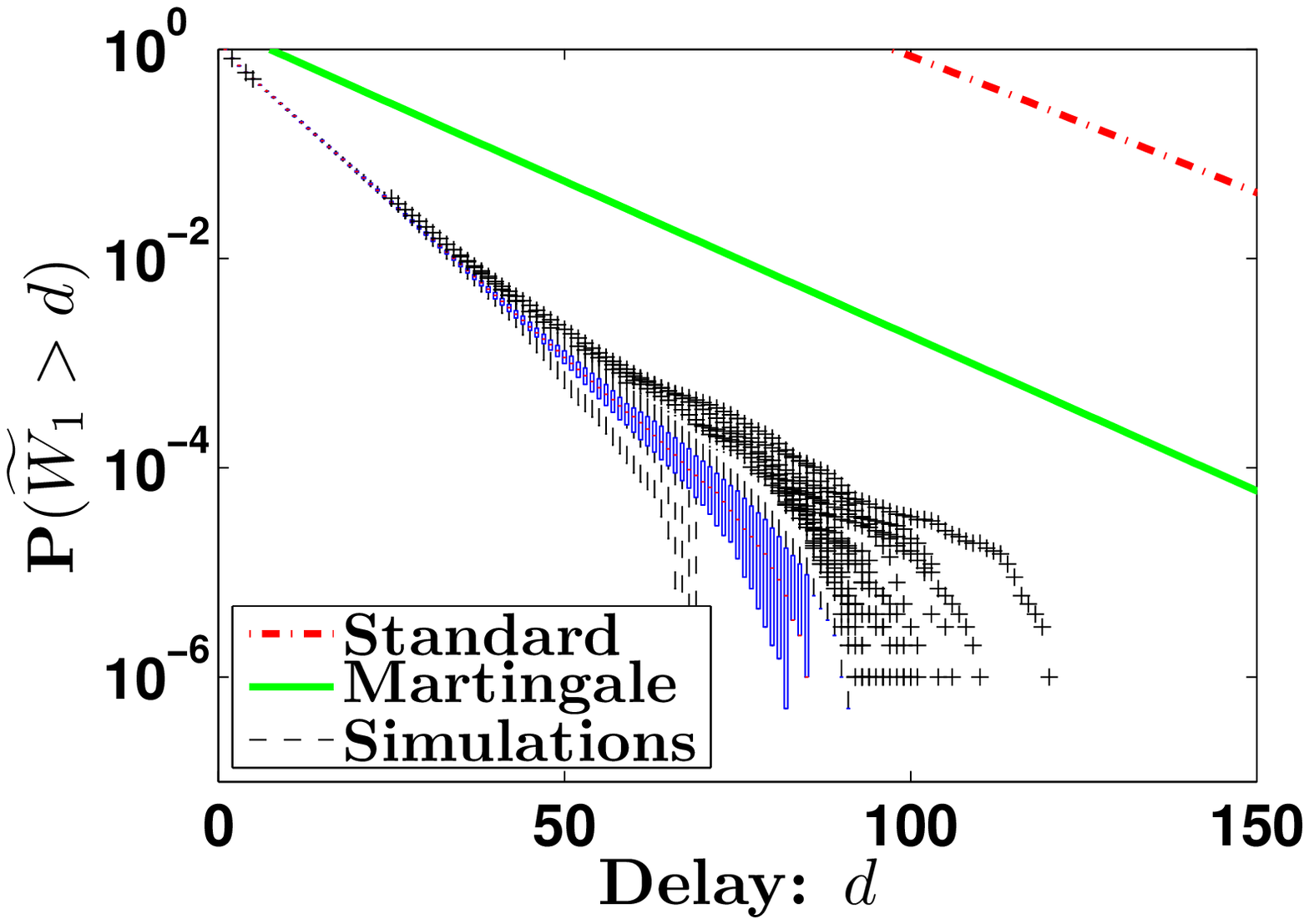}
\\
{\footnotesize (b) $\rho=0.90$} ($n_1=n_2=5$)}
 \caption{GPS delay
bounds} \label{fig:wfq}
\end{figure}

The positive side of our exposition so far (i.e., the Martingale
bounds fix the very loose Standard bounds) is disturbed, however,
in the case of GPS scheduling (with $\phi_1=\phi_2=.5$). Indeed,
let us refer to Figure~\ref{fig:wfq} showing the corresponding
numerical comparisons (recall the Martingale and Standard bounds
from Eqs.~(\ref{eq:gpsdb}) and (\ref{eq:gpsdbStandard})); we
mention that the packet-level simulator implements WFQ scheduling,
i.e., the packetized version of GPS (see Keshav~\cite{keshav97},
pp. 238-242). Unlike in the FIFO, SP, and EDF cases, the
Martingale bounds are now very loose. This indicates that, in
addition to the Union Bound bounding step, the `proprietary' SNC
bounding step from Eq.~(\ref{eq:genb}) can also be very loose. In
other words, the service process for GPS scheduling from
Eq.~(\ref{eq:sgps}) needs to be significantly improved.

\subsubsection{Asymptotic comparisons}
Here we illustrate the bounds in a many sources asymptotic regime:
we fix the target delay bound $d=5$ and plot the per-flow
violation probabilities $\P\left(\widetilde{W}_1>d\right)$ by
increasing the total number of flows $n$. For visualization
purposes we restrict to FIFO, SP, and one case of EDF.

\begin{figure}[h]
\vspace{-0cm}
\shortstack{\hspace{-0.165cm}
\includegraphics[width=0.515\linewidth,keepaspectratio]{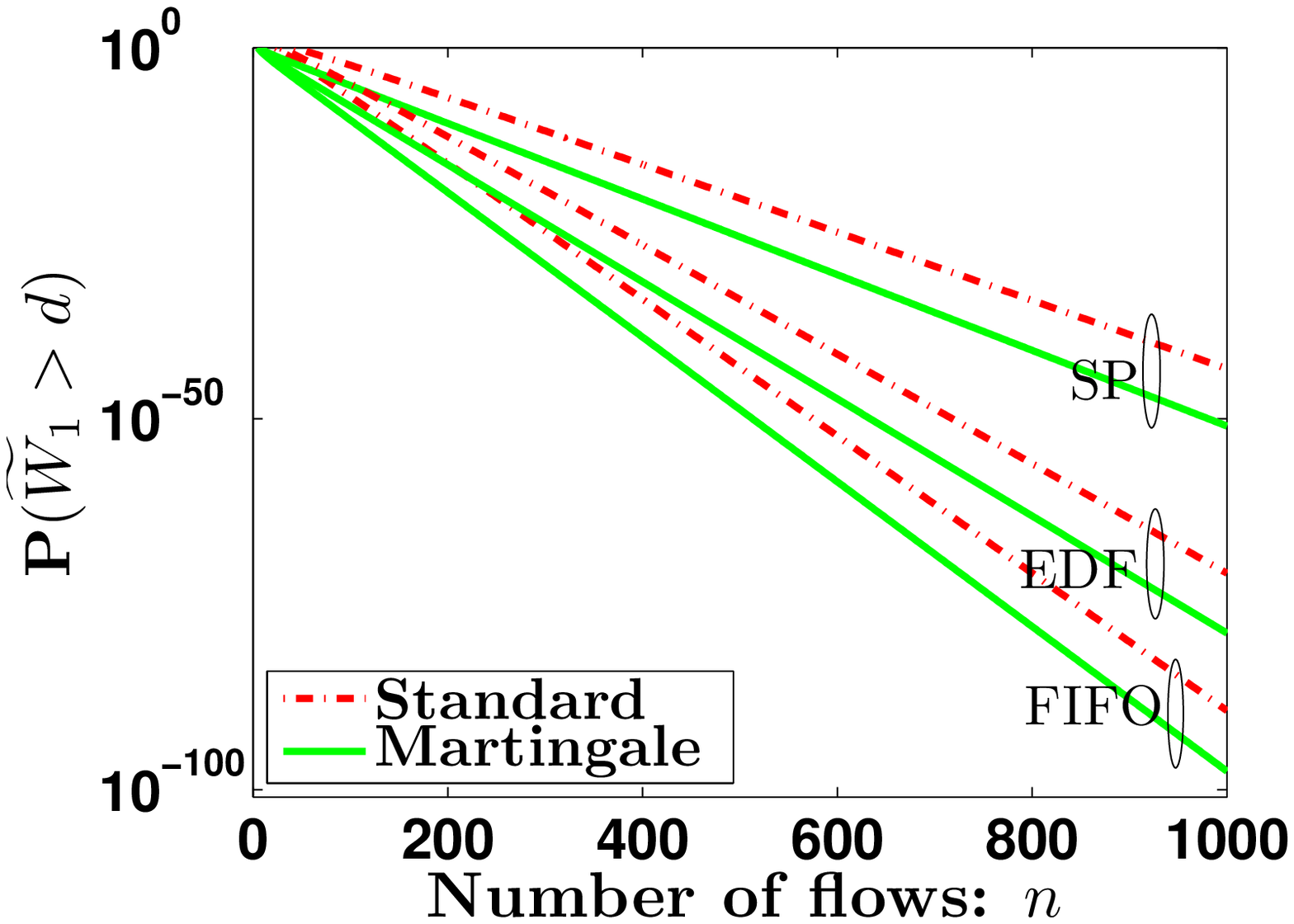}
\\
{\footnotesize (a) $\rho=75\%$} } \shortstack{\hspace{-0.41cm}
\includegraphics[width=0.517\linewidth,keepaspectratio]{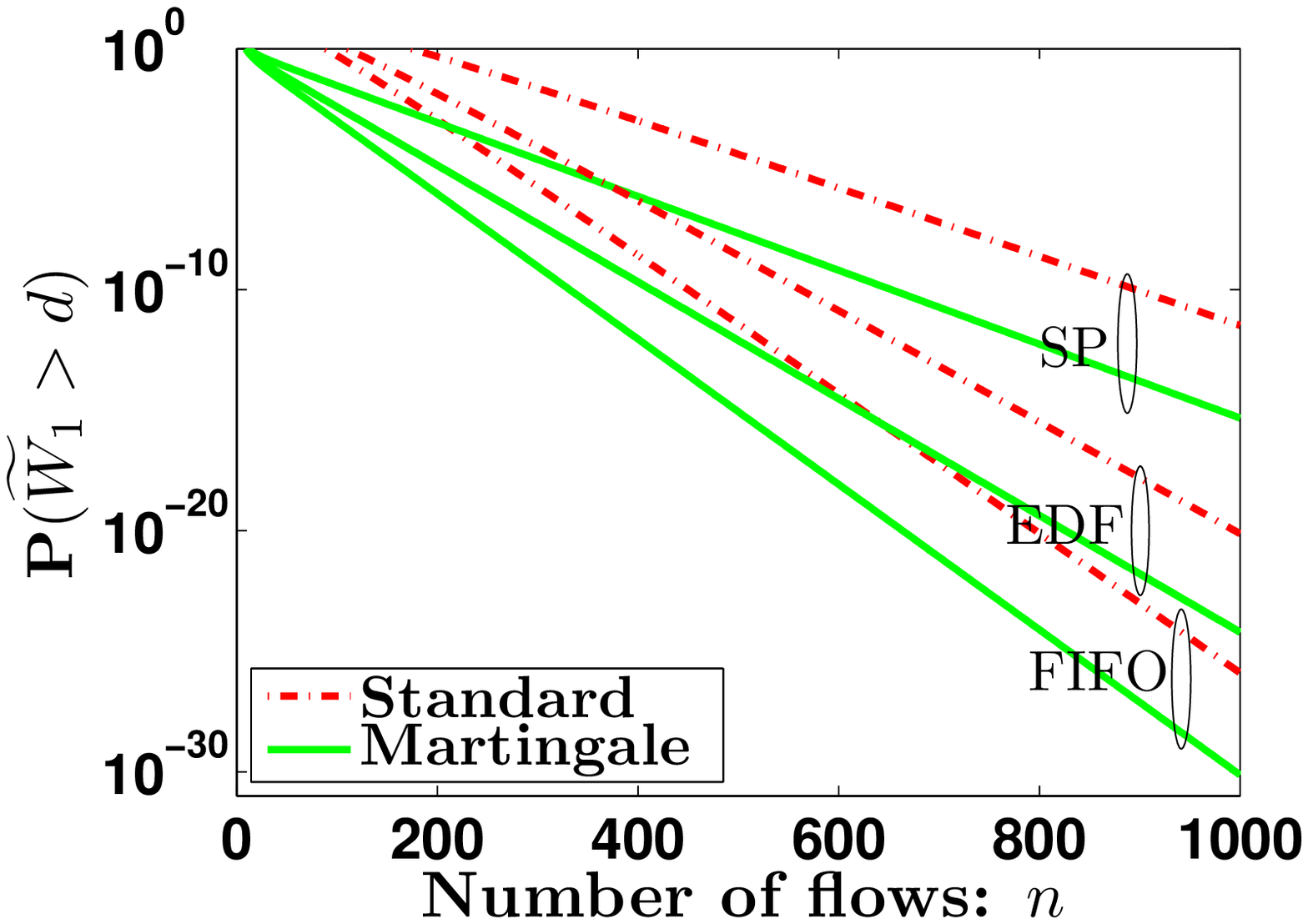}
\\
{\footnotesize (b) $\rho=90\%$}}
 \caption{Violation probabilities for a delay $d=5$ as a function of the total number of flows $n=n_1+n_2$ ($n_1=n_2$, for EDF: $d_1^*=3$ and $d_2^*=1$)} \label{fig:asc}
\end{figure}

The important observation from Figure~\ref{fig:asc} is that, for
all settings (e.g., FIFO, $\rho=75\%$), the Standard and
Martingale bounds diverge. In other words the discrepancy between
the Standard and Martingale bounds grows arbitrarily large (e.g.,
by a factor of $10^{9}$ at $n=10^{3}$). This observation supports
the scaling laws from Table~\ref{tb:slaws}. In the light of this
rather pessimistic evidence concerning the Standard bounds, the
immediate question is whether they can still have any practical
relevance; next we clarify this concern.

\subsubsection{\hspace{-0.1cm}The Standard bounds can (sometimes) be
useful}\label{sec:useful} Here we show that, despite the poor
scaling in multiplexing regimes, the Standard bounds can be
practically relevant. To this end, we consider the following
connection admission control problem: given the server capacity
$C$, the target (per-flow) delay bound $d$, and a violation
probability $\eps$, determine the maximum number of flows $n$ to
meet all the three constraints.

\begin{figure}[h]
\vspace{-0cm}
\shortstack{\hspace{-0.165cm}
\includegraphics[width=0.515\linewidth,keepaspectratio]{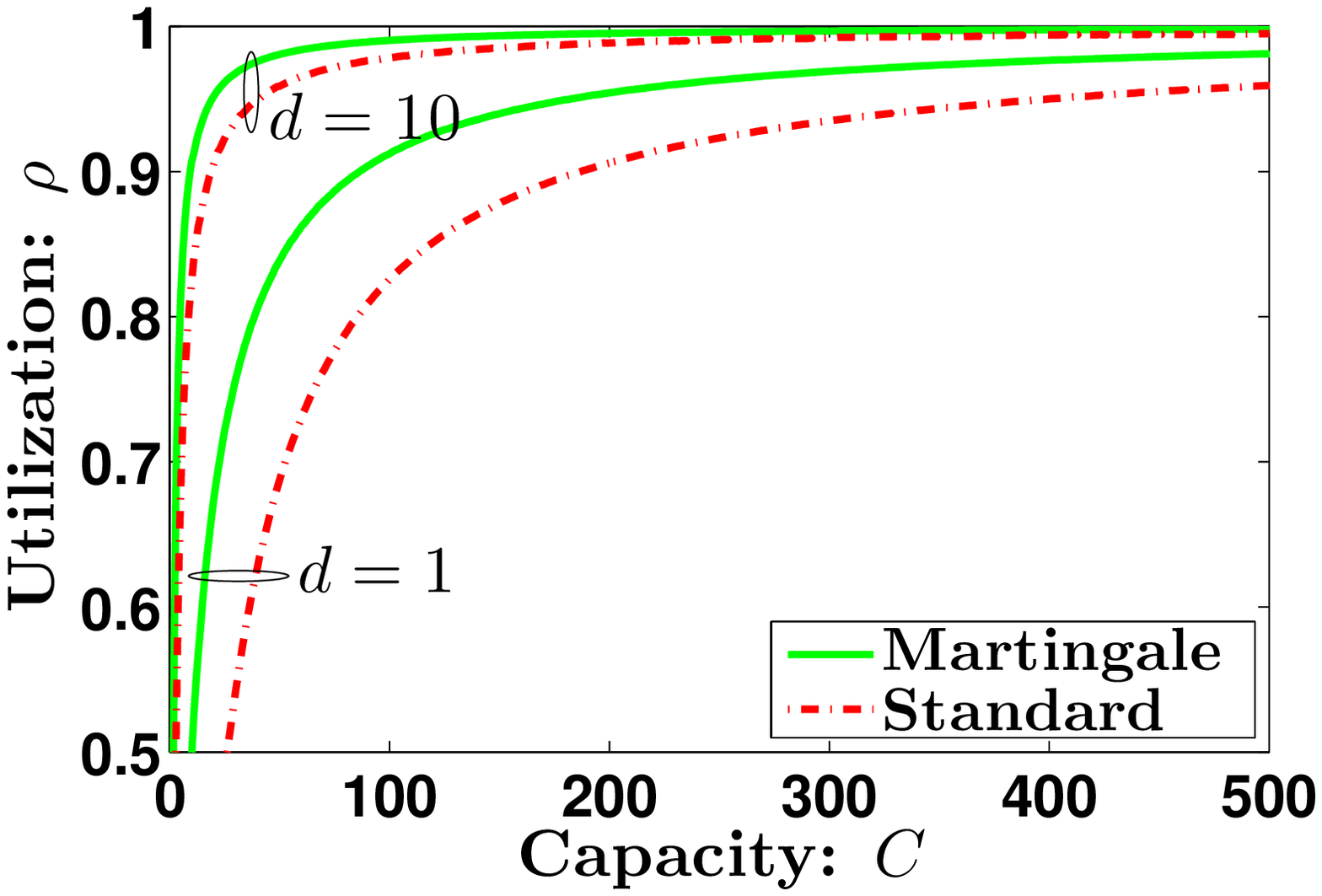}
\\
{\footnotesize (a) FIFO, $\eps=10^{-3}$} }
\shortstack{\hspace{-0.475cm}
\includegraphics[width=0.515\linewidth,keepaspectratio]{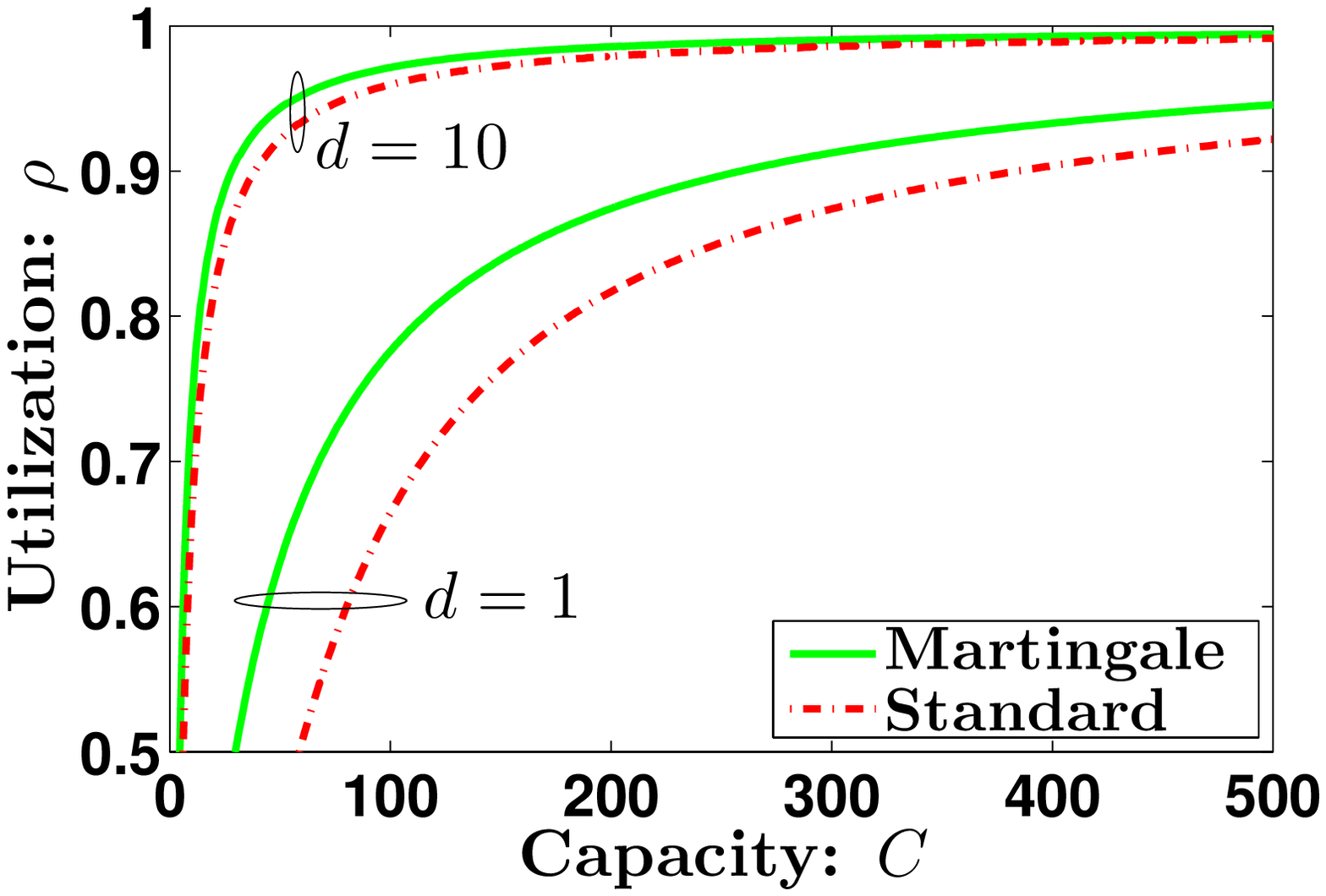}
\\
{\footnotesize (b) FIFO, $\eps=10^{-9}$}}
\shortstack{\hspace{-0.165cm}
\includegraphics[width=0.515\linewidth,keepaspectratio]{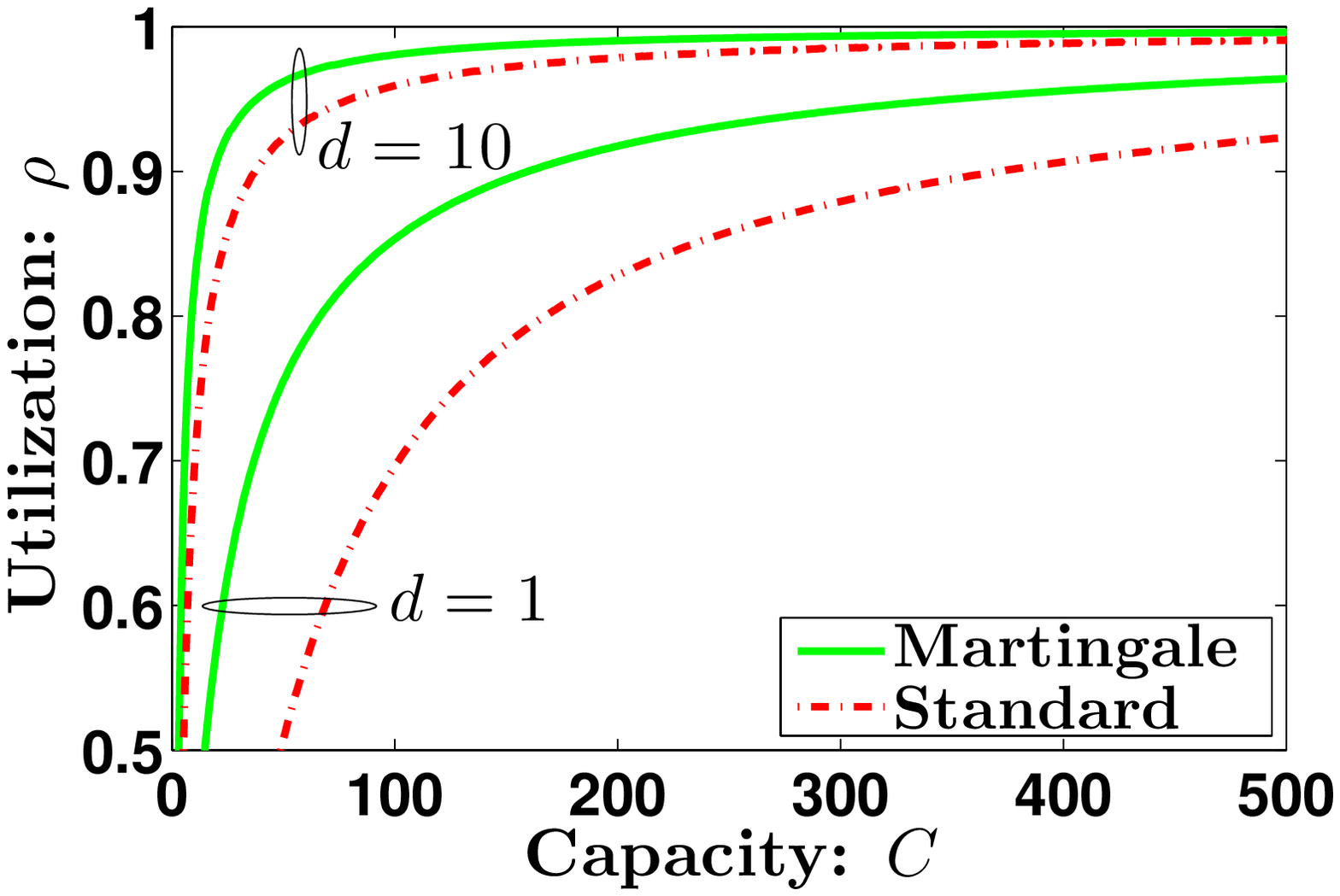}
\\
{\footnotesize (c) SP, $\eps=10^{-3}$} }
\shortstack{\hspace{-0.475cm}
\includegraphics[width=0.515\linewidth,keepaspectratio]{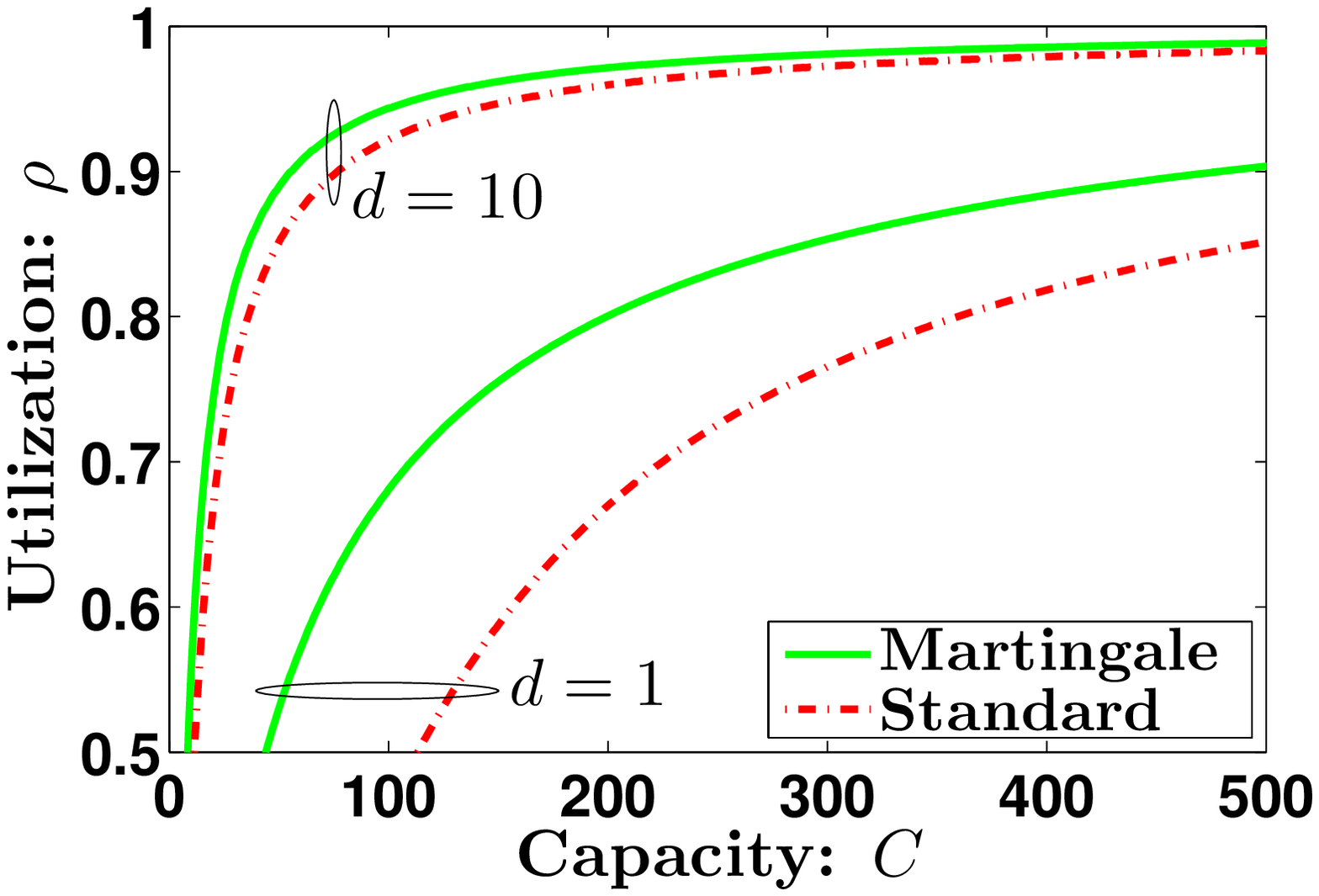}
\\
{\footnotesize (d) SP, $\eps=10^{-9}$}}
 \caption{Achievable utilization as a function of the capacity $C$ ($n_1=n_2$)} \label{fig:cac}
\end{figure}

Figure~\ref{fig:cac} shows the achievable utilization (i.e.,
$\rho=\frac{npP}{C}$) as a function of the capacity $C$ for FIFO
and SP, two delay targets $d=1$ and $d=10$, and two violation
probabilities $\eps=10^{-3}$ and $10^{-9}$. The important
observation is that once $d$ and $\eps$ are both fixed, then by
increasing the capacity $C$ (i.e., `making room for sufficient
statistical multiplexing to kick in') the Martingale and Standard
bounds converge to one, albeit at different rates.

The convergence for both FIFO and SP is a manifestation of the
assertion from the literature that ``scheduling has only a limited
impact" (e.g., on admission control) (see, e.g.,
Li~\et~\cite{LiBuLi07}); such an assertion, however, does not
generally hold (see, e.g., Figures~\ref{fig:fifo}, \ref{fig:sp},
and \ref{fig:asc}).

We conclude that the Standard bounds can (sometimes) be nearly
optimal and thus be practically relevant. This remark is typically
presented in the literature in the form: the admissible region
based on the Standard bounds converges to the admissible region
based on averages~(see, e.g., Boorstyn~\et~\cite{Boorstyn2000b} or
Li~\et~\cite{LiBuLi07}). Variations of the underlying scaling
regime, enabling such conclusions, are often adopted in the
literature in order to expose the Standard bounds in a favorable
light.

\section{Conclusions}\label{sec:conclusions}
In this paper, we have put our finger in a wound of the stochastic
network calculus: the lingering issue of the tightness of the SNC
bounds. To some degree, this issue had been evaded by the SNC
literature for some time although it is a, if not the, crucial
one. In fact, we demonstrated that the typical (Standard SNC) way
of calculating performance bounds results in loose delay bounds
for several scheduling disciplines (FIFO, SP, EDF, and GPS) as
well as for various multiplexing regimes. This becomes
particularly obvious when comparing the (Standard SNC) analytical
results to simulation results, where discrepancies up to many
orders of magnitude can be observed. So, we strongly confirm the
often rumored conjecture about SNC's looseness.

Yet, the paper does not stop at these bad news, but in an attempt
to understand the problems of Standard SNC, which mainly lie in
not properly accounting for the correlation structure of the
arrival processes (by coarse usage of the Union bound), we find a
new way to calculate performance bounds using the SNC framework
based on martingale techniques. Here, SNC still serves as the
``master method", yet the Union bound is substituted by the usage
of martingale inequalities, to make a long story short. Comparing
the new Martingale SNC bounds to the simulation results shows that
they are remarkably close in most cases, which rehabilitates the
SNC as a general framework for performance analysis. So, the SNC
can arguably still be regarded as a valuable methodology with the
caveat that it has to be used with the right probabilistic
techniques in order not to arrive at practically irrelevant
results. As usual, there is a ``but" to such a general statement:
this is the issue about GPS for which even the Martingale bounds
stay loose (though again improving by orders of magnitude over the
Standard bounds). This may be indeed a case where SNC as master
method fails, because it might be wrong to separate the service
process derivation from the arrivals of the flow under
investigation. On the other hand, this is very speculative and it
may still be possible to find a service process that tightly
captures the GPS characteristics. We leave this interesting open
issue for future work.

Other related challenges include extensions to general Markov
arrival processes (e.g., by using the general result
from~\cite{Palmowski96}), self-similar arrival processes, and the
multi-node case (e.g., by accounting for the martingale
representation from~\cite{chang96}).

\bibliographystyle{abbrv}
\bibliography{../../../stat}
\newpage
\appendix

\subsection{Extension to General Markov Fluid Processes}
Here we present the generalization of Theorem~\ref{th:tspb} to the
case when the arrival processes $A_1(t)$ and $A_2(t)$ are general
Markov fluid processes.

Consider the queueing model from Figure~\ref{fig:qsystem}, in
which $A_1(t)$ and $A_2(t)$  are two cumulative Markov fluid
processes served at constant-rate $C$. Each process $A_k(t)$ is
modulated by a reversible Markov process $Z_k(t)$ with $n_k+1$
states, generator
$\mathbf{Q}_k=\left(q_{k,i,j}\right)_{i,j=0,\dots,n_k}$,
equilibrium distribution
$\mathbf{\pi}_k=\left(\pi_{k,0},\dots,\pi_{k,n_k}\right)$, arrival
rates $\mathbf{r}_k=\left(r_{k,0},\dots,r_{k,n_k}\right)$, and
increment process $a_k(s)=r_{k,Z_k(s)}$ $\forall s\geq0$. We
remark that if $Z_k(t)$ were not reversible, then one could
consider as input the corresponding reversed processes for the
sake of expressing Reich's equation as
\begin{equation*}
Q=\sup_{t\geq0}\left\{A_1(t)+A_2(t)-Ct\right\}~.
\end{equation*}

For each arrival process $A_k(t)$ we consider the generalized
eigenvalue problem
\begin{equation}
\mathbf{Q}_k
\mathbf{h}_k=-\gamma_{k}\mathbf{u}_k\mathbf{h}_k~k=1,2~,\label{eq:eigprob}
\end{equation}
where $\mathbf{Q}_k$ are the generators, $\mathbf{u}_k$ are
diagonal matrices with
$\left(u_{k,0},u_{k,1},\dots,u_{k,n_k}\right)$ on the diagonal,
and where \begin{equation*}u_{k,j}=r_{k,j}-C_k\end{equation*} are
the instantaneous queueing drifts for $j=0,1,\dots,n_k$. Here,
$C_1$ and $C_2$ are positive values such that $C_1+C_2=C$, and can
be regarded as the per-class allocated capacity.

Assuming the per-class stability conditions
\begin{equation}
\sum_{j=0}^{n_k}\pi_{k,j}u_{k,j}<0,~k=1,2~,\label{eq:stabcond}
\end{equation}
Lemma~5.1 from~\cite{Palmowski96} guarantees the existence of real
generalized eigenvalues $-\gamma_k$ (as the ones with the biggest
negative real parts) and also of the generalized eigenvectors
$\mathbf{h}_k=\left(h_{k,0},h_{k,1},\dots,h_{k,n_k}\right)^{\textrm{T}}$
with positive coordinates. Thus, $\gamma_k>0$ for $k=1,2$.

\begin{theorem}{({\sc{A General Sample-Path Bound}})}\label{th:tspb}
Consider the single-node queueing scenario from
Figure~\ref{fig:qsystem} and the solutions for the generalized
eigenvalue problems from Eq.~(\ref{eq:eigprob}). Then the
following sample-path bound holds for all $0\leq u\leq t$ and
$\sigma$
\begin{eqnarray}
&&\hspace{-1cm}\P\left(\sup_{0\leq s<
t-u}\left\{A_1(s,t-u)+A_2(s,t)-C(t-s)\right\}>\sigma\right)\leq\inf_{0\leq\gamma\leq\min\gamma_k}\inf_{C_1+C_2=C}Ke^{-\gamma\left(C_1
u+\sigma\right)}~,\label{eq:tspb}
\end{eqnarray}
where
$K=\frac{\sum_{i,j}\pi_{1,i}\pi_{2,j}h_{1,i}^{\frac{\gamma}{\gamma_1}}h_{2,j}^{\frac{\gamma}{\gamma_2}}}{\min_{u_{1,i}+u_{2,j}\geq0}h_{1,i}^{\frac{\gamma}{\gamma_1}}h_{2,j}^{\frac{\gamma}{\gamma_2}}}$
whereas the condition $C_1+C_2=C$ is subject to the stability
conditions from Eq.~(\ref{eq:stabcond}).
\end{theorem}

\medskip

Let us make several observations the two infimum operators. The
parameter $\gamma$ in the former infimum reconciles the different
burstiness of the two not-necessarily homogeneous flows $A_1(t)$
and $A_2(t)$, loosely expressed through the exponential decay
factor. The extreme optimal value
$\gamma=\min\left\{\gamma_1,\gamma_2\right\}$ is attained for
$\sigma\rightarrow\infty$; in turn, an optimization after $\gamma$
is necessary in finite regimes of $\sigma$. In turn, due to the
implicit expression of $K$ in terms of the (generalized)
eigenvectors from Eq.~(\ref{eq:eigprob}), which depend on $C_1$
and $C_2$, the optimal values for the former infimum are not
apparent and hence numerical optimizations must be invoked.

The theorem generalizes Theorem~\ref{th:tspb} to the case of
general and not-necessarily homogenous Markov fluid processes
(recall that Theorem~\ref{th:tspb} is restricted to multiplexed
homogenous Markov-modulated On-Off processes). The theorem also
generalizes the seminal result of Palmowski and
Rolski~\cite{Palmowski96} (see Proposition~5.1 therein),
restricted to $A_2(t)=0$; more details on the extent of our
generalization will be provided after the proof.

\medskip

\proof Fix $u\geq0$ and $\sigma$. Since the two arrival processes
are reversible, we can rewrite the probability from
Eq.~(\ref{eq:tspb}), by shifting the time origin, as
\begin{eqnarray}
&&\hspace{-1cm}\P\left(\sup_{t>u}\left\{A_1(u,t)+A_2(t)-Ct\right\}>\sigma\right)\notag\\
&&=\P\Bigg(\sup_{t>u}\left\{A_1(u,t)+A_2(u,t)-C(t-u)\right\}+A_2(u)-C_2u>C_1u+\sigma\Bigg)~.\label{eq:revspb}
\end{eqnarray}

Given the last probability event we construct the stopping time
\begin{eqnarray}
&&T:=\inf\Bigg\{t>u:A_1(u,t)+A_2(u,t)-C(t-u)+A_2(u)-C_2u>C_1u+\sigma\Bigg\}~.\label{eq:stot}
\end{eqnarray}
For the rest of the proof we bound $\P\left(T<\infty\right)$,
which exactly characterizes the probability from
Eq.~(\ref{eq:revspb}).

Let $\P_{i,j}$ denote the underlying probability measure
conditioned on $Z_1(u)=i$ and $Z_2(0)=j$, for $0\leq i\leq n_1$
and $0\leq j\leq n_2$. Next we define the following two processes
\begin{eqnarray*}
\widetilde{M}_{1,t}&:=&\frac{h_{1,Z_1(t)}}{h_{1,i}}e^{-\int_u^t\frac{\left(\mathbf{Q}_1\mathbf{h}_1\right)_{Z_1(s)}}{h_{1,Z_1(s)}}ds}~\forall
t\geq u~\textrm{and}\\
\widetilde{M}_{2,t}&:=&\frac{h_{2,Z_2(t)}}{h_{2,j}}e^{-\int_0^t\frac{\left(\mathbf{Q}_2\mathbf{h}_2\right)_{Z_2(s)}}{h_{2,Z_2(s)}}ds}~\forall
t\geq0~.
\end{eqnarray*}
$M_1(t)$ and $M_2(t)$ are supermartingales with respect to (wrt)
$\P_{i,j}$ and the natural filtration (see Ethier and
Kurtz~\cite{EK86}, p. 175). Considering the solution of the
generalized eigenvalue problem from Eq.~(\ref{eq:eigprob}), we can
rewrite
\begin{eqnarray*}
\widetilde{M}_{1,t}&=&\frac{h_{1,Z_1(t)}}{h_{1,i}}e^{\gamma_1\int_u^t
u_{1,Z_1(s)}ds}~\forall
t\geq u~\textrm{and}\\
\widetilde{M}_{2,t}&=&\frac{h_{2,Z_2(t)}}{h_{2,j}}e^{\gamma_2\int_0^t
u_{2,Z_2(s)}ds}~\forall t\geq0~.
\end{eqnarray*}

For $0\leq\gamma\leq\min\left\{\gamma_1,\gamma_2\right\}$ we
consider the transformations
\begin{equation*}
M_{k,t}=\widetilde{M}_{k,t}^{\frac{\gamma}{\gamma_k}}~k=1,2~.
\end{equation*}
Denoting by ${\mathcal{F}}_{k,s}$ the natural filtrations of
$M_{k,t}$ we can write for $0\leq s\leq t$
\begin{eqnarray*}
E\left[M_{k,t}\mid{\mathcal{F}}_{k,s}\right]&=&
E\left[\widetilde{M}_{k,t}^{\frac{\gamma}{\gamma_k}}\mid{\mathcal{F}}_{k,s}\right]\leq
E\left[\widetilde{M}_{k,t}\mid{\mathcal{F}}_{k,s}\right]^{\frac{\gamma}{\gamma_k}}\\&\leq&\widetilde{M}_{k,s}^{\frac{\gamma}{\gamma_k}}=M_{k,s}~,
\end{eqnarray*}
where the first inequality is due to Jensen's inequality (applied
to the concave function $\psi(x)=x^\frac{\gamma}{\gamma_k}$ for
$x\geq0$) and the second inequality is due to the supermartingale
property of $\widetilde{M}_{k,t}$. Therefore, the new processes
$M_{k,t}$ are also supermartingales; we point out that there
construction is motivated by the need of having the same decay,
i.e., $\gamma$, in the corresponding exponentials.

Next we invoke Lemmas~\ref{lm:pim} and~\ref{lm:os} from the
Appendix and obtain that the process
\begin{equation*}
M_t:=\left\{\begin{array}{lcc}M_{2,t}&,&t\leq
u\\M_{1,t}M_{2,t}&,&t> u\end{array}\right.
\end{equation*}
is also a supermartingale (note that $M_{1,u}=1$ by definition).
It can be explicitly written as
\begin{equation*}
M_t=\left\{\begin{array}{lcc}\left(\frac{h_{2,Z_2(t)}}{h_{2,j}}\right)^{\frac{\gamma}{\gamma_2}}e^{\gamma\left(A_2(t)-C_2
t\right)},t\leq
u\\\left(\frac{h_{1,Z_1(t)}}{h_{1,i}}\right)^{\frac{\gamma}{\gamma_1}}\left(\frac{h_{2,Z_2(t)}}{h_{2,j}}\right)^{\frac{\gamma}{\gamma_2}}e^{\gamma\left(A_1(u,t)+A_2(t)-C
t\right)},t> u\end{array}\right.
\end{equation*}

Referring now to the stopping time $T$, which may be unbounded, we
construct the bounded stopping times $T\wedge v$ for all $v\in\N$.
For these times, the Optional Sampling theorem (see
Theorem~\ref{th:ost} in the Appendix) yields
\begin{eqnarray*}
E_{i,j}\left[M_0\right]=E_{i,j}\left[M_{T\wedge v}\right]~,
\end{eqnarray*}
for all $v\in N$, where the expectations are taken wrt the
underlying probability measure $\P_{i,j}$. Moreover, from the
definition of $T$ as an infimum over a set, it holds for $v\geq0$
that
\begin{equation}
\left(u_{1,Z_1(T)}+u_{2,Z_2(T)}\right)I_{\left\{T\leq
v\right\}}\geq0~,\label{eq:condreach}
\end{equation}
where $I_{\left\{\cdot\right\}}$ denotes the indicator function.
Using now that $E_{i,j}\left[M_0\right]=1$ we obtain for $v>u$
\begin{eqnarray*}
1&\geq&E_{i,j}\left[M_{T\wedge v}I_{\left\{T\leq v\right\}}\right]\\
&\geq&
\min\left(\frac{h_{1,l_1}}{h_{1,i}}\right)^{\frac{\gamma}{\gamma_1}}\left(\frac{h_{2,l_2}}{h_{2,j}}\right)^{\frac{\gamma}{\gamma_2}}e^{\gamma\left(
C_1 u+\sigma\right)}\P_{i,j}\left(T\leq v\right)~,
\end{eqnarray*}
where the `$\min$' operator is taken over the set
$\left\{(l_1,l_2):u_{1,l_1}+u_{2,l_2}\geq0\right\}$ according to
Eq.~(\ref{eq:condreach}).

Finally, by deconditioning on $i$ and $j$ (recall that $Z_1(u)$
and $Z_2(0)$ are in steady-state by construction) we obtain
\begin{equation*}
\P\left(T\leq v\right)\leq Ke^{-\gamma\left( C_1
u+\sigma\right)}~.
\end{equation*}
Letting $v\rightarrow\infty$ and optimizing after $\gamma$
completes the proof.~\hfill $\Box$

\subsubsection{Relationship to State-of-the-Art Bound} Here we show
that the bound from Theorem~\ref{th:tspb} improves the
state-of-the-art result from~\cite{Palmowski96}. In the particular
case of an arrival flow consisting of $n$ multiplexed
Markov-modulated On-Off processes, the improvement is of the order
${\mathcal{O}}\left(K^n\right)$ for some constant $0<K<1$.

The state-of-the-art bound from~\cite{Palmowski96} concerns the
distribution of the (stationary) queue size occupancy for a single
Markov fluid process served at rate $C$. To fit this scenario in
Theorem~\ref{th:tspb}, we let $A_2(t)=0$ and $u=0$, and for
convenience we drop the index in the parameters of the remaining
process $A_1(t)$. Our bound states that
\begin{equation}
\P\left(Q>\sigma\right)\leq\frac{\sum_i\pi_{i}h_{i}}{\min_{u_i\geq0}h_i}e^{-\gamma\sigma}\label{eq:ourbound}
\end{equation}
where $\gamma$ and
$\mathbf{h}=\left(h_0,h_1,\dots,h_n\right)^{\textrm{T}}$ are the
solution of the generalized eigenvalue problem from
Eq.~(\ref{eq:eigprob}). In turn, the bound from~\cite{Palmowski96}
states that
\begin{equation}
\P\left(Q>\sigma\right)\leq\frac{\sum_i\pi_{i}h_{i}}{\min_i
h_i}e^{-\gamma\sigma}~.\label{eq:prbound}
\end{equation}

The bound from Eq.~(\ref{eq:ourbound}) is clearly tighter than the
bound from Eq.~(\ref{eq:prbound}); see the additional constraint
on the `$\min$' operator in Eq.~(\ref{eq:ourbound}). Next we give
the order of the improvement when the process $A(t)$ is a
superposition of $n$ Markov-modulated On-Off processes. Each
sub-process is modulated by a Markov process $Z(t)$ with two
states, denoted by `On' and `Off', and which communicate at rates
$\lambda$ and $\mu$. While in the `On' state, each sub-process
generates data units at a constant rate $P$. In this case,
according to Eq.~(\ref{eq:ourbound}), the queue size distribution
is bounded by
\begin{equation*}
\P\left(Q>\sigma\right)\leq K^ne^{-\gamma\sigma}~,
\end{equation*}
for some $0<K<1$; see Theorem~1 in~\cite{CiPoSc13}, which recovers
Theorem~2.1 from~\cite{Palmowski96}. In turn, the pre-factor from
Eq.~(\ref{eq:prbound}) satisfies $\frac{\sum_i\pi_{i}h_{i}}{\min_i
h_i}\geq1$, whence the ${\mathcal{O}}\left(K^n\right)$ improvement
of the bound from Eq.~(\ref{eq:ourbound}) over the one from
Eq.~(\ref{eq:prbound}).

\subsubsection{Relationship to Effective Bandwidth} Here we
establish a fundamental relationship between the decay rate
$\gamma$ and the effective bandwidth.

We consider a single arrival process $A_1(t)$ for which we drop
the index, i.e., $A(t)$. The effective bandwidth of $A(t)$ is
defined for $\theta>0$ as
\begin{equation*}
\alpha\left(\theta,t\right):=\frac{1}{t\theta}\log
E\left[e^{\theta A(t)}\right]~,
\end{equation*}
and let
$\alpha_{\theta}:=\lim_{t\rightarrow\infty}\alpha\left(\theta,t\right)$
(see Kelly~\cite{Kelly96}); with abuse of notation
$\alpha_{\theta}$ will be called the effective bandwidth of
$A(t)$.

\begin{lemma}{({\sc{$\gamma$ vs. $\alpha_{\theta}$}})}\label{cor:gammavsalpha}
Let the scenario from Theorem~\ref{th:tspb}, with a single arrival
process $A(t)$ having effective bandwidth
$\alpha\left(\theta,t\right)$. If $\alpha_{\theta}$ is
differentiable then
\begin{equation}
\alpha_{\gamma}=C~.
\end{equation}
\end{lemma}

\proof From the construction of $\gamma$ from the generalized
eigenvalue problem from Eq.~(\ref{eq:eigprob}), for which we drop
the indexes, we have that
\begin{equation}
\left(\mathbf{Q}+\gamma\mathbf{u}\right)\mathbf{h}=0~.\label{eq:gepv2}
\end{equation}

Let the diagonal matrix $\mathbf{V}$ with
$\left(r_0\gamma,r_1\gamma,\dots,r_n\gamma\right)$ on the
diagonal, and construct the matrix
\begin{equation*}
{\mathbf{Q}}_{\gamma}:=\mathbf{Q}+\mathbf{V}~.
\end{equation*}
Then it holds that
\begin{equation}
\mathbf{Q}_{\gamma}\mathbf{x}=\alpha_{\gamma}\gamma\mathbf{I}\mathbf{x}\label{eq:epv2}
\end{equation}
where $\alpha_{\gamma}\gamma$ is the spectral radius of
$\mathbf{Q}_{\gamma}$ and $\mathbf{x}$ is the corresponding
(positive) eigenvector (see~Kesidis~\et~, Sec.
3,~\cite{KeWaCh93}).

Let us now observe that
\begin{equation*}
\mathbf{Q}+\gamma\mathbf{u}=\mathbf{Q}_{\delta}-C\gamma\mathbf{I}~.
\end{equation*}
Combining with Eqs.~(\ref{eq:gepv2}) and (\ref{eq:epv2}) we obtain
that
\begin{equation*}
0=\left(\mathbf{Q}_{\gamma}-\alpha_{\gamma}\gamma\mathbf{I}\right)\mathbf{x}=\left(\mathbf{Q}_{\gamma}-C\gamma\mathbf{I}\right)\mathbf{h}~.
\end{equation*}
Therefore, $C\gamma$ is an eigenvalue for the eigenvalue problem
from Eq.~(\ref{eq:epv2}) and thus
\begin{equation*}
\alpha_{\gamma}\geq C~,\label{eq:gore}
\end{equation*}
since by construction $\alpha_{\gamma}\gamma$ is the corresponding
spectral radius.

To show the converse, i.e., $\alpha{\gamma}\leq C$, consider the
exact asymptotic decay of the distribution of the queue occupancy
(of $A(t)$ when fed at a queue with capacity $C$), i.e.,
\begin{equation*}
\lim_{\sigma\rightarrow\infty}\frac{1}{\sigma}\P\left(Q>\sigma\right)=-\theta^*~,
\end{equation*}
where $\alpha_{\theta^*}=C$ (see Kelly,
Eq.~(3.21),~\cite{Kelly96}). In other words, $\theta^*$ is the
exact asymptotic decay rate. As Theorem~\ref{th:tspb} predicts
$\gamma$ as a decay rate, in terms of an upper bound, it follows
that $\gamma\leq\theta^*$. Finally, since $\alpha_{\theta}$ is
increasing in $\theta$ (see Chang,~p.241,~\cite{Book-Chang}), it
follows that
\begin{equation*}
\alpha_{\gamma}\leq\alpha_{\theta^*}=C~,
\end{equation*}
completing thus the proof that $\alpha_{\gamma}=C$.~\hfill $\Box$

\subsection{Review of Martingale Results}

Here we summarize some definitions and results related to
martingales, which are needed in the paper.

Consider a (continuous-time) stochastic process $(X_t)_{t\geq0}$
defined on some joint probability space
$(\Omega,{\mathcal{F}},\P)$. A filtration is a family
$\left\{\mathcal{F}_t:t\geq0\right\}$ of sub-$\sigma$-fields of
${\mathcal{F}}$ such that
${\mathcal{F}}_s\subseteq{\mathcal{F}}_t$ for all $0\leq s\leq t$.
We are particularly interested in the natural filtration generated
by $X_t$, i.e., ${\mathcal{F}}^X_t:=\sigma\left(X_s:s\leq
t\right)$.

\begin{definition}{({\sc{Stopping Time}})}\label{def:st}
A stopping time, wrt a filtration ${\mathcal{F}}_t$, is a
non-negative random variable $T$ such that
\begin{equation*}\left\{T\leq t\right\}\in{\mathcal{F}}_t~\forall t\geq0~.
\end{equation*}
\end{definition}

In this paper we are particularly interested in the first passage
time
\begin{equation*}
T=\inf_{t\geq0}\left\{X_t\geq x\right\}~,
\end{equation*}
for some non-negative $x$. To avoid technical considerations
related to conditions under which $T$ is a stopping time, we
assume throughout that the sample-paths $X_t(\omega)$ are
right-continuous $\forall\omega\in\Omega$; this assumption is
implicitly fulfilled by the definition of the arrival process from
Eq.~(\ref{eq:mmoosf}).

\begin{definition}{({\sc{Martingale}})}\label{def:mart}
A continuous time process $X_t$ is a martingale wrt the natural
filtration ${\mathcal{F}}^X_t$ if
\begin{enumerate}
\item $E\left[\mid X_t\mid\right]<\infty$ $\forall t\geq0$ and
\item $E\left[X_t\mid{\mathcal{F}}_s\right]=X_s$ $\forall 0\leq
s\leq t$~.
\end{enumerate}
\end{definition}

The next three results are needed in the proof of the main theorem
in the paper.

\begin{lemma}{({\sc{Optional Switching}})}\label{lm:os}
Consider that $X_t$ and $Y_t$ are martingales wrt
${\mathcal{F}}:={\mathcal{F}}_t^{X,Y}$, and assume that $X_u=Y_u$
for some $u\geq0$. Then the process
\begin{equation*}
Z_t=\left\{\begin{array}{ccc}X_t&,&~\textrm{if}~~t<u\\Y_t&,&~\textrm{if}~~t\geq
u\end{array}\right.
\end{equation*}
is a martingale wrt ${\mathcal{F}}$.
\end{lemma}
For the corresponding result in discrete time see Grimmett and
Stirzaker~\cite{Grimmett01}, p. 488.

\proof According to the martingale definition, the non-trivial
property to prove is
\begin{equation*}
E\left[Z_t\mid{\mathcal{F}}_s\right]=Z_s~,
\end{equation*}
for $t\geq u$ and $s<u$. Indeed, we have according to the tower
property of conditional expectation
\begin{eqnarray*}
E\left[Z_t\mid{\mathcal{F}}_s\right]&=&E\left[E\left[Z_t\mid{\mathcal{F}}_u\right]\mid{\mathcal{F}}_s\right]\\
&=&E\left[Y_u\mid{\mathcal{F}}_s\right]=E\left[X_u\mid{\mathcal{F}}_s\right]\\
&=&Z_s~,
\end{eqnarray*}
which completes the proof. ~\hfill $\Box$

\begin{lemma}{({\sc{Product of Independent Martingales}})}\cite{Cherny06}\label{lm:pim}
Consider that $X_t$ and $Y_t$ are independent martingales wrt
${\mathcal{F}}:={\mathcal{F}}_t^{X,Y}$. Then $X_tY_t$ is a
martingale wrt ${\mathcal{F}}$.
\end{lemma}

\begin{theorem}{({\sc{Optional Sampling Theorem}})}(see~\cite{EK86}, p. 61)\label{th:ost}
If $X_t$ is a right-continuous martingale and $T$ is a finite
stopping time wrt ${\mathcal{F}}_t^{X}$, then
\begin{equation*}
E\left[X_T\right]=E\left[X_0\right]~.
\end{equation*}
\end{theorem}

\end{document}